\documentclass{aa}  

\usepackage{graphicx}
\usepackage{txfonts}
\usepackage{xcolor}
\usepackage[breaklinks=true]{hyperref} 
\hypersetup{
	colorlinks = true, % colours links instead of ungly boxes
	urlcolor = red,  % color for external hyperlinks
	linkcolor = blue, % color of internal links
	citecolor = blue % color of citations
}
\usepackage{mathastext}

\newcommand{\mum}{$\mu$m}

\newcommand{\oiit}{[\ion{O}{ii}]$\lambda3726,3729$} 
\newcommand{\hbl}{H$\beta$$\lambda4861$} 
\newcommand{\oiiibl}{[\ion{O}{iii}]$\lambda4959$} 
\newcommand{\oiiirl}{[\ion{O}{iii}]$\lambda5007$} 
\newcommand{\oibl}{[\ion{O}{i}]$\lambda6300$} 
\newcommand{\oirl}{[\ion{O}{i}]$\lambda6363$} 
\newcommand{\hal}{H$\alpha$$\lambda6562$} 
\newcommand{\niibl}{[\ion{N}{ii}]$\lambda6548$} 
\newcommand{\niirl}{[\ion{N}{ii}]$\lambda6584$} 
\newcommand{\niit}{[\ion{N}{ii}]$\lambda6548, 6584$} 
\newcommand{\siibl}{[\ion{S}{ii}]$\lambda6716$} 
\newcommand{\siirl}{[\ion{S}{ii}]$\lambda6731$} 
\newcommand{\siit}{[\ion{S}{ii}]$\lambda6716,6731$} 
\newcommand{\hb}{H$\beta$} 
\newcommand{\oiii}{[\ion{O}{iii}]} 
\newcommand{\oi}{[\ion{O}{i}]} 
\newcommand{\ha}{H$\alpha$} 
\newcommand{\nii}{[\ion{N}{ii}]} 
\newcommand{\sii}{[\ion{S}{ii}]} 
\newcommand{\fevii}{[\ion{Fe}{vii}]$\lambda4989$}
\newcommand{\hii}{\ion{H}{ii}} 

\newcommand{\neiil}{[\ion{Ne}{ii}]$12.8\mu$m} 
\newcommand{\siiil}{[\ion{S}{iii}]$18.7\mu$m} 
\newcommand{\siiilred}{[\ion{S}{iii}]$33.5\mu$m} 
\newcommand{\sivl}{[\ion{S}{iv}]$10.5\mu$m} 
\newcommand{\neiiil}{[\ion{Ne}{iii}]$15.7\mu$m} 
\newcommand{\nevl}{[\ion{Ne}{v}]$14.3\mu$m} 
\newcommand{\nevrl}{[\ion{Ne}{v}]$24.3\mu$m} 
\newcommand{\oivl}{[\ion{O}{iv}]$25.9\mu$m} 

\newcommand{\neii}{[\ion{Ne}{ii}]} 
\newcommand{\siii}{[\ion{S}{iii}]} 
\newcommand{\siv}{[\ion{S}{iv}]} 
\newcommand{\nev}{[\ion{Ne}{v}]} 
\newcommand{\oiv}{[\ion{O}{iv}]} 
\newcommand{\neiii}{[\ion{Ne}{iii}]}

\newcommand{\logU}{log($\langle$U$\rangle$)}
\newcommand{\ionpar}{$\langle$U$\rangle$}

\newcommand{\orcid}[1]{\href{https://orcid.org/#1}{\includegraphics[scale=0.3]{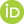}\,}}

\begin{document}

   \title{Optical and mid-infrared line emission in nearby Seyfert galaxies}

   %\subtitle{}
   
   %\titlerunning{}

   \author{A.~Feltre 
          \inst{\orcid{0000-0001-6865-2871} 1} \thanks{E-mail: anna.feltre@inaf.it, feltre.anna@gmail.com}
          \and
          C.~Gruppioni \inst{\orcid{0000-0002-5836-4056} 1}
          \and
          L.~Marchetti \inst{\orcid{0000-0003-3948-7621} 2,3}
          \and 
          A.~Mahoro \inst{\orcid{0000-0002-6518-781X} 4,2}
          \and
          F.~Salvestrini \inst{\orcid{0000-0003-4751-7421} 5}
          \and
          M.~Mignoli \inst{\orcid{0000-0002-9087-2835} 1}
          \and
          L.~Bisigello\inst{\orcid{0000-0003-0492-4924} 6,7}
          \and
          F.~Calura\inst{\orcid{0000-0002-6175-0871} 1}
          \and
          S.~Charlot\inst{\orcid{0000-0003-3458-2275} 8}
          \and
          J.~Chevallard\inst{\orcid{0000-0002-7636-0534} 9}
          \and
          E.~Romero-Colmenero \inst{\orcid{0000-0003-0607-1136} 4, 10}
           \and
           E.~Curtis-Lake\inst{\orcid{0000-0002-9551-0534} 11}
           \and
           I.~Delvecchio\inst{\orcid{0000-0001-8706-2252} 12}
           \and
           O. L.~Dors\inst{\orcid{0000-0003-4782-1570} 13}
           \and
           M.~Hirschmann\inst{\orcid{0000-0002-3301-3321} 14,15}
           \and
          T.~Jarrett\inst{\orcid{0000-0002-4939-734X} 2}
          \and
          S.~Marchesi\inst{\orcid{0000-0001-5544-0749} 1,16}
          \and
          M.~E.~Moloko\inst{\orcid{0000-0001-5519-0620} 2}
          \and 
          A.~Plat\inst{\orcid{0000-0003-0390-0656} 17,8}
          \and
          F.~Pozzi\inst{\orcid{0000-0003-1006-924X} 18}
          \and
          R.~Sefako \inst{\orcid{0000-0003-3904-6754} 4}
          \and
          A.~Traina\inst{\orcid{0000-0003-1006-924X} 1,19}
          \and 
          M.~Vaccari \inst{\orcid{0000-0002-6748-0577} 19,20 ,3}
          \and
          P.~V\"ais\"anen \inst{\orcid{0000-0001-7673-4850} 4}
          \and
          L.~Vallini\inst{\orcid{0000-0002-3258-3672} 21}
          \and
          A.~Vidal-Garc\'ia\inst{\orcid{0000-0001-6798-9202} 22,23}
         \and
          C.~Vignali \inst{\orcid{0000-0002-8853-9611} 18}
          }

   \institute{INAF  -  Osservatorio  di  Astrofisica  e  Scienza  dello  Spazio  di  Bologna,  Via  P.  Gobetti  93/3,  40129  Bologna,  Italy
                \and
                Department of Astronomy, University of Cape Town, Private Bag X3, Rondebosch 7701, South Africa
                \and 
                INAF - Istituto di Radioastronomia, via Gobetti 101, I-40129 Bologna, Italy
                \and
                South African Astronomical Observatory, P.O. Box 9, Observatory 7935, Cape Town, South Africa
                \and
                INAF - Osservatorio Astrofisico di Arcetri, Largo E. Fermi 5, 50125, Firenze, Italy
                \and
                Dipartimento di fisica e astronomia, Università di Padova, Vicolo dell'Osservatorio 3, I-35122, Padova, Italy
                \and
                INAF - Osservatorio astronomico di Padova, Vicolo dell'Osservatorio 5, I-35122, Padova, Italy
                \and
                Sorbonne Universit\'e, CNRS, UMR 7095, Institut d’Astrophysique de Paris, 98 bis bd Arago, 75014 Paris, France
                \and
                Sub-department of Astrophysics, Department of Physics, University of Oxford, Denys Wilkinson Building, Keble Road, Oxford OX1 3RH, UK 
                \and
                Southern African Large Telescope, P.O. Box 9, Observatory, 7935, South Africa 
                \and
                Centre for Astrophysics Research, Department of Physics, Astronomy and Mathematics, University of Hertfordshire, Hatfield, AL10 9AB, UK
                \and
                INAF - Osservatorio Astronomico di Brera, via Brera 28, I-20121, Milano, Italy \& via Bianchi 46, I-23807 Merate, Italy
                \and
                UNIVAP - Universidade do Vale do Paraíba. Av. Shishima Hifumi, 2911, CEP: 12244-000 São José dos Campos, SP, Brazil
                \and
                Institut de Physique, Laboratoire d'astrophysique, École Polytechnique Fédérale de Lausanne (EPFL), Observatoire de Sauverny, Chemin Pegasi 51, CH-1290 Versoix, Switzerland
                \and
                INAF -Osservatorio Astronomico di Trieste, via G.B. Tiepolo 11, I-34143 Trieste, Italy              
                \and
                Department of Physics and Astronomy, Clemson University, Kinard Lab of Physics, Clemson, SC 29634, USA
                \and
                Steward Observatory, University of Arizona, 933 N Cherry Ave., Tucson, AZ 85721, USA
                \and
                Dipartimento di Fisica e Astronomia, Università degli Studi di Bologna, via Gobetti 93/2, 40129, Bologna, Italy
                \and
                The Inter-University Institute for Data Intensive Astronomy, Department of Astronomy, University of Cape Town, Private Bag X3, Rondebosch 7701, South Africa
                \and
                nter-University Institute for Data Intensive Astronomy, Department of Physics and Astronomy, University of the Western Cape, 7535 Bellville, Cape Town, South Africa
                \and
                Scuola Normale Superiore, Piazza dei Cavalieri 7, I-56126 Pisa, Italy
                \and
                Observatorio Astronómico Nacional, C/ Alfonso XII 3, 28014 Madrid, Spain
                \and
                Ecole Normale Supérieure, CNRS, UMR 8023, Université PSL, Sorbonne Université, Université de Paris, F-75005, Paris, France
                }

   \date{Received Month X, XXXX; accepted Month X, XXXX}

% \abstract{}{}{}{}{} 
% 5 {} token are mandatory
 
  \abstract{
Line ratio diagnostics provide valuable clues as to the source of ionizing radiation in galaxies with intense black hole accretion and starbursting events, such as local Seyfert galaxies or galaxies at the peak of their star formation history. We aim to provide a reference joint optical and mid-IR line ratio analysis for studying active galactic nucleus (AGN) identification via line-ratio diagnostics and testing predictions from photoionization models. We first obtained homogenous optical spectra with the Southern Africa Large Telescope for 42 Seyfert galaxies with available \textit{Spitzer}/IRS spectroscopy, along with X-ray to mid-IR multiband data. After confirming the power of the main optical (\oiiirl) and mid-IR (\nevl, \oivl, \neiiil) emission lines in tracing AGN activity, we explored diagrams based on ratios of optical and mid-IR lines by exploiting photoionization models of different ionizing sources (AGN, star formation, and shocks). We find that pure AGN photoionization models are good at reproducing observations of Seyfert galaxies with an AGN fractional contribution to the mid-IR ($5-40$ \mum) continuum emission larger than 50 per cent. For targets with a lower AGN contribution, even assuming a hard ionizing field from the central accretion disk ($F_{\nu} \propto \nu^{\alpha}$, with $\alpha$ $\approx -0.9$), these same models do not fully reproduce the observed mid-IR line ratios. Mid-IR line ratios such as \nevl/\neiil,  \oivl/\neiil,\ and \neiiil/\neiil\ show a dependence on the AGN fractional contribution to the mid-IR, unlike optical line ratios. An additional source of ionization, either from star formation or radiative shocks, can help explain the observations in the mid-IR. While mid-IR line ratios are good tracers of the AGN activity versus star formation, among the combinations of optical and mid-IR diagnostics in line-ratio diagrams, only those involving the \oi/\ha\ ratio are promising diagnostics for simultaneously unraveling the relative roles of AGN, star formation, and shocks. A proper identification of the dominant source of ionizing photons would require the exploitation of analysis tools based on advanced statistical techniques as well as spatially resolved data. }

% \keywords{Galaxies: Seyfert -- Galaxies: ISM -- Infrared: galaxies -- ISM: lines and bands}

   \maketitle
%
%-------------------------------------------------------------------

\section{Introduction}

Over the years, specific combinations of intensity ratios of emission lines have been proposed as a means to identify gas ionized by the radiation from black hole accretion in obscured (i.e., Type 2) active galactic nuclei (AGN), where the presence of dust prevents a direct view of the accretion disk \citep{Rowan-Robinson1977, Antonucci1993, UrryPadovani1995}. The most commonly used diagnostics for disentangling gas ionized by the AGN radiation from stellar-driven ionization by O and B stars are based on the ratios of strong optical lines (such as \ha, \hb, \oiit, \oiiirl, \niit, \siit, and \oibl), first proposed by \citet*[][BPT]{BPT} and \cite{Veilleux1987}. Thereafter, optical emission-line ratios have been routinely exploited to search for AGN in statistical samples of galaxies \citep[e.g.,][]{Kauffmann2003, Groves2006b, Juneau2014}. More recently, integral field spectroscopy from, for example, the Multi Unit Spectroscopic Explorer \citep[MUSE][]{Bacon2010} on the Very Large Telescope (VLT) and surveys such as the Calar Alto Legacy Integral Field Area Survey \citep[CALIFA][]{Sanchez2012}, the Mapping Nearby Galaxies at Apache Point Observatory \citep[MaNGA][]{Bundy2015}, and the Sydney – Australian Astronomical
Observatory Multi-Object Integral Field Spectrograph (SAMI) Galaxy Survey \citep{Allen2015} have enabled studies of the spatial extent of the AGN impact on the interstellar medium (ISM) of their host galaxies \citep[e.g.,][]{Dagostino2019, Mingozzi2019, Lacerda2020, Deconto2022}.

 The emission lines measured in galaxy spectra are often interpreted by means of photoionization models, developed using standard photoionization codes such as \textsc{cloudy} \citep{Ferland1993, Ferland1998, Ferland2013, Ferland2017} and \textsc{mappings} \citep{Binette1985, SutherlandDopita1993, Dopita2013, Sutherland2018}, which enables an analysis of the features ascribable to different ionizing sources in galaxies as well as of the physical conditions of the ionized gas within them. While optical lines provide valuable constraints on the physical properties of the ionizing sources and the physical conditions of the ISM of galaxies in the local Universe, detailed rest-frame optical spectroscopy at z$\gtrsim1-2$ is currently confined to small samples of galaxies and AGN. In addition, the separation between AGN and star formation activity through optical diagnostic diagrams becomes less clear as the physical conditions of the ISM in galaxies evolve with cosmic time. \cite{Kewley2013} show that the radiation field in the ISM of  $z \: > \:1.5$ star-forming galaxies is harder than in local objects, calling into question the current use of optical diagnostic diagrams at high redshifts. Moreover, \cite{Hirschmann2017} and \cite{Curti2022} find that the position of galaxies in the \oiiirl/\hb\ versus \niirl/\ha\ BPT diagram depends on the physical properties of the galaxies themselves, such as the stellar mass, star formation rate (SFR), gas-phase metallicity, and dust content. Finally, as the metallicity decreases with increasing redshift, the areas of the BPT diagram populated by sources ionized by AGN or young stars overlap \citep{Groves2006c, Feltre2016, Hirschmann2019}.
This called for new diagnostic diagrams to be used as alternative or alongside the standard optical ones. 

Since current optical and near-IR ground-based spectrographs can probe the redshifted UV emission from star-forming galaxies and AGN at $z \gtrapprox1-2$, several works have investigated the power of UV features to act as diagnostics of the properties of the stellar populations and the ISM in galaxies \citep[e.g.,][]{Gutkin2016, Vidal-Garcia2017, Byler2017, Byler2020} and to disentangle different ionizing sources, such as young stars, shocks, and AGN. UV line ratios have proven useful in distinguishing AGN-ionized gas from shock-ionized gas \citep[e.g., ][]{Villar-Martin1997, Allen1998, Best2000, Humphrey2008} and discriminating between nuclear activity and star formation \citep[][]{Feltre2016, Nakajima2018, Dors2018, Hirschmann2019, Plat2019, Nakajima2022}.

One of the main uncertainties affecting optical and UV lines is that they are sensitive to dust attenuation, motivating the exploration other features that do not require resorting to dust corrections, for example those in the IR domain. The rest-frame mid-IR regime exclusively offers probes of the emission from different low-to-high ionization states of several elements, such as argon, neon, sulfur, and oxygen, arising from the most obscured (dusty) regions within galaxies, which are out of reach in optical and near-IR surveys. Specifically, IR features such as [\ion{Ar}{ii}]6.9\mum, [\ion{Ar}{iii}]9.0\mum, [\ion{Ar}{v}]8, 13.1\mum, \neiil, \neiiil, \nevl, \nevrl, [\ion{Ne}{vi}]7.7\mum, \siiil\, \sivl, [\ion{O}{i}]63, 145\mum, [\ion{O}{iii}]52, 88\mum, \oivl, [\ion{N}{ii}]122\mum, and [\ion{C}{ii}]158\mum\  directly trace the primary source of ionizing radiation that cannot be observed in the far-UV/X-ray in the presence of Galactic and intrinsic absorption. Different combinations of ratios of low-to-high IR ionization lines, explored with photoionization models, have been found to help identify and quantify the contribution from different ionizing sources, such as AGN, star formation, and shocks to line emission, and help in the study of the physical conditions of the ISM, such as metal content \citep[e.g.,][]{Fernandez-Ontiveros2016, Fernandez-Ontiveros2017, Riffel2013}.

Mid- and far-IR emission lines are currently available for local sources, mainly from {\it Spitzer}$/$InfraRed Spectrograph (IRS) \citep{Houck2004}, {\it Herschel}$/$Photodetector Array Camera and Spectrometer  (PACS). \citep{Poglitsch2010}, and \,{\it Herschel}$/$Spectral and Photometric Imaging Receiver (SPIRE) \citep{Griffin2010, Naylor2010} spectrometers. Yet, at $z \gtrapprox 4$, the Atacama Large Millimeter/submillimeter Array (ALMA) targets rest-frame far-IR lines of high redshift galaxies and quasars \citep{Walter2009, Maiolino2015, Inoue2016, Harikane2020, Carniani2020, Decarli2020, Bethermin2020, Venemans2020, LeFevre2020, Bouwens2022}. Mid-IR spectroscopy is a very promising avenue for expanding current observational and theoretical studies in the coming years. Data from the Mid-Infrared Instrument \citep[MIRI][]{Rieke2015} of the recently launched \textit{James Webb} Space Telescope (JWST) will enable unprecedented detailed spatially resolved studies of mid-IR lines in local targets. Moreover, mid-IR line-ratio diagnostics have been proposed as powerful tools for the identification of local dwarf AGN \citep{Richardson2022} and elusive AGN \citep[i.e., AGN undetectable via commonly employed methods due to obscurations or contamination from star formation;][]{Satyapal2021} in JWST/MIRI observations. In addition, future IR missions, for example the PRobe far-Infrared Mission for Astrophysics\footnote{\url{https://workshop.ipac.caltech.edu/farirprobe/system/media_files/binaries/2/original/prima_factsheet_v1.2.pdf?1645918445}}\citep[PRIMA;] []{Glenn2021}, will observe mid- and far-IR features up to and beyond the peak of the cosmic star formation history ($1.5 < z  < 3$). 

Usually, line ratios in different wavelength regimes are analyzed separately as rest-frame optical/UV and mid/far-IR information is not always available for the same targets; there are a few exceptions when focusing on star-forming galaxies rather than sources spectroscopically classified as AGN \citep[see, for instance,][]{Abel2008, Dors2013, Dors2016}. On the theoretical side, a few works have investigated UV, optical, and/or IR emission line models of AGN \citep{Groves2004, Groves2006a, Spinoglio2015, Fernandez-Ontiveros2016}, shocks \citep{Allen2008}, and star formation \citep{Inami2013,Fernandez-Ontiveros2016}. Detailed theoretical modeling that combines rest-frame optical and mid-IR spectroscopy for AGN has not been performed yet, and few studies have addressed this for star-forming galaxies \cite[e.g.,][]{Perez-Montero2009}. To overcome this lack, we collected homogeneous optical and mid-IR spectra for a sample of local Seyfert galaxies with the main aim of investigating the role of the different physical processes (AGN, star formation, and shocks) through the study of the observed ratios of lines of different ionization levels and the exploitation of photoionization models. Calibrating models in the optical and mid-IR simultaneously will be instrumental for interpreting new JWST/MIRI data and for the design of future mid-IR follow-up spectroscopic observations of $z\gtrsim2-3$ optically selected galaxies.

For this purpose, we targeted AGN in the local Universe, which are ideal laboratories for detailed spectroscopic studies of galaxies whose emission is powered by black hole accretion and intense star formation. These composite objects are local analogs of sources at  $z\approx1-3$ that dominate the peaks of the total IR luminosity density and SFR density \citep{Gruppioni2013}. For objects detected by  {\em Herschel} at redshifts $z>1-2$, direct spectroscopic follow-up of standard rest-frame optical diagnostics is a challenge, as they fall outside the wavelength range accessible to most current spectrographs. In this context, the local IR galaxies we focus on here represent an ideal reference sample for understanding the physics at play not only in such objects at low redshifts, but also in their {\em Herschel} high-$z$ analogs.
More specifically, the Seyfert galaxies from the {\it InfraRed Astronomical Satellite} (IRS) 12 $\mu$m local Seyfert galaxy sample \citep[12MGS;][]{Rush1993} are the best local benchmarks for studying the physical conditions in composite objects (where AGN and star formation coexist) because of the availability of spectrophotometric data across the whole electromagnetic spectrum. For example, \citet[][hereafter G16]{Gruppioni2016} collected multiband data, X-ray to submillimeter AF{ Units should be spelled out in full when not following a numeral.},  for 76 Seyfert galaxies from the 12MGS. While \textit{Spitzer}/IRS high-resolution spectra are currently available for all these targets, no homogenous optical spectra had been collected. For this reason, we designed the ``SALT Spectroscopic Survey of IR 12MGS Seyfert Galaxies.''  The goal of this observational campaign was to collect homogeneous optical spectra with the Southern Africa Large Telescope (SALT) over the wavelength range 3600 to 7500 \AA\ for a significant number of local Seyferts in the G16 sample (i.e., 43 out of the 76 sources in the southern hemisphere).

The sample and the SALT observations, along with the multiband data of our targets, are described in Sect. \ref{sec:data_and_sample}. The analysis of the optical spectra from SALT is presented in Sect. \ref{sec:spectral_analysis}. We then confront the \oiiirl\ flux with other AGN tracers in Sect. \ref{sec:lacc_tracers}. We explore mid-IR diagnostic diagrams and compare observations with predictions from photoionization models in Sects. \ref{sec:mir} and \ref{sec:model_comparison}, respectively. The results from this work and some caveats of the analysis are discussed in Sect. \ref{sec:discussion}, followed by the conclusions in Sect. \ref{sec:conclusions}. 
Throughout the paper, we use an initial mass function from \cite{chabrier2003}, with lower and upper mass cutoffs of 0.01 and 100 M$_{\odot}$, respectively, and we adopt the cosmological parameters from \cite{Planck2020}, ($\Omega_{\rm M}$, $\Omega_{\Lambda}$, H$_{0}$) = (0.315, 0.685, 67.4 kms$^{-1}$Mpc$^{-1}$).

%--------------------------------------------------------------------
\section{Data and sample}\label{sec:data_and_sample}

We selected the 43 sources from the 76 of the G16 sample that were observable with SALT. 
These are part of the 12 \mum\ galaxy sample (12MGS) from \cite{Rush1993}, which comprises 893 galaxies with IRAS 12\mum\ flux larger than 0.22 Jy, 118 of which, including all our targets, were classified as AGN/Seyfert from existing catalogs of active galaxies \citep{Spinoglio1989, Hewitt1991, Veron1991}. According to the most recent spectral classifications, our sample comprises galaxies of different types, specifically seven Seyfert 1 (Sy1), nine Seyfert 2 (Sy2), twenty-two intermediate ($1.2-1.9$) Seyfert (int-Sy), and five non-Seyfert (non-Sy) galaxies that are either \ion{H}{ii} regions or low ionization nuclear emission-line regions (LINERs). This classification is taken from \cite{BrightmanNandra2011b} and \cite{Malkan2017} and is based on optical line-ratio diagnostic diagrams, luminosities of the \oiiirl\ line, the 12\mum\ continuum  and X-ray spectral properties. The class of intermediate Seyferts includes also sources classified as Sy2 in previous works but with evidence for hidden broad line regions, as detailed in \cite{Tommasin2008,Tommasin2010}, which show broad lines either in their IR spectra (Sect. \ref{sec:ir}) or polarized light spectra. We adopt this classification throughout the present work (column 5 of Table \ref{table:prop}).

The G16 (see their Sect. 2.1) sample, selected to have homogeneous {\it Spitzer}/IRS spectroscopy, can boast richness of multiband data, from X-ray to far-IR. G16 performed a detailed spectral energy distribution (SED) decomposition analysis of the multiband photometry, including contributions from stars, AGN and dust, inferring the main physical properties of the galaxies in their sample (e.g., AGN bolometric luminosity, SFR, total IR luminosity, stellar mass and fractional contribution of the AGN to the total emission). In Table \ref{table:prop} we report the AGN bolometric luminosity, L$_{\rm bol}$(IR), computed from the 1-1000 $\mu$m rest frame luminosity of the best-fitting AGN template, and the fractional contribution of the AGN to the total continuum emission in the 5-40 \mum\ spectral range, f$_{\rm AGN}$, for the Seyfert galaxies out of the 76 from G16 studied in this work. According to the definition of the main sequence of star-forming galaxies from \cite{Renzini2015} and the classification in four regions in the stellar mass--SFR plane on the basis of the distance from it \citep[][]{Bluck2020}, our targets are either main sequence or starburst galaxies, with stellar masses in the range $ 9.0 <log(M_{\star} / M_{\odot})< 11.5$ and SFR in the range $0.49<M_{\odot}/yr<155$. For the other properties inferred from the SED decomposition analysis that are not considered in this work, we refer to Table 3 of G16. Another feature is that {\it Spitzer}/IRS high- and low- resolution spectra \citep{Tommasin2008,Tommasin2010, Wu2009} are available for all our targets (Sect. \ref{sec:ir}), enabling a joint analysis of optical and mid-IR spectral properties in the same targets.

%\begin{table*}
\begin{sidewaystable*}
\caption{Main information and properties of the 33 12MGS targets studied in this work}        
\label{table:prop}     
\centering                          
\begin{tabular}{l c c c c c c c c c }       
\hline\hline                 
Name  & R.A. & Dec. & z &  Sy type & f$_{\rm AGN}$ & log(L$_{\rm bol}$(IR)) & log(L$_{\rm bol}$([\ion{Ne}{v}])) & log(L$_{\rm bol}$([\ion{O}{iv}]) )& X-ray \\   
  & (J2000) & (J2000) &  &   &5-40\mum & log(erg/s) & log(erg/s) & log(erg/s) & Reference\\    
\hline                        
 CGCG381-051   & 23:48:41.3 & 02:14:21 & 0.03067 & non-Sy& $0.01\pm0.42$ &  $43.06\pm0.27$  &  $<43.80$  &  40.6 & 1\\    
 ESO033-G002  & 04:55:59.6 & -75:32:27 & 0.0181 & Sy2 & $0.61\pm0.02$ & $44.34\pm0.01$& $44.47\pm0.07$& 43.02 & 1 \\
 ESO141-G055   & 19:21:14.3 & $-$58:40:13 & 0.03710 &  Sy1 & $0.86 \pm 0.02$ & $45.01\pm0.02$ & $44.70\pm0.01$ & 43.93 & 3  \\
 ESO362-G018 & 05:19:35.5 &  $-$32:39:30 & 0.0124 & int-Sy & $0.48\pm0.6$ &$44.27\pm0.30$ & $43.71\pm0.10$ & 42.73 & 4 \\
IC4329A & 13:49:19.3 & $-$30:18:34 & 0.01605 & int-Sy &  $0.88 \pm0.02$ & $45.11\pm0.00$ & $45.24\pm0.11$ & 43.85 & 5 \\
IC5063 & 20:52:02.0 & $-$57:04:09 & 0.01135  & int-Sy & $0.46\pm0.05$  & $44.85\pm0.30$ & $44.76\pm0.04$ & 42.87 & 1 \\
IRASF01475-0740 & 01:50:02.7 & $-$07:25:48 & 0.01766 & int-Sy&  $0.25\pm0.60$ & $43.55\pm0.05$ & $44.38\pm0.20$ & 41.85 & 1 \\
IRASF03450+0055 & 03:47:40.2 & +01:05:14 & 0.031 & Sy1 &  $0.83\pm0.02$  &$44.77\pm0.01$ & $<44.25$ &... & ... \\
IRASF04385-0828 & 04:40:54.9 & $-$08:22:22 & 0.0151 & int-Sy & $0.71\pm0.03$ &  $44.82\pm0.02$ & $43.70\pm0.20$ & 43.55 & 1 \\
IRASF05189-2524 & 05:21:01.4 & $-$25:21:45 & 0.0426 & Sy2 & $0.12\pm0.04$ & $45.38\pm0.03$ & $46.03\pm0.27$ & 43.39 & 1 \\ 
IRASF15480-0344 & 15:50:41.5 & $-$03:53:18 & 0.03030 & int-Sy &  $0.52\pm0.10$ & $45.09\pm0.30$  & $45.04\pm0.06$ & 43.28 & 1 \\
MCG-02-33-034 & 12:52:12.4 & $-$13:24:54 & 0.0146 7 & Sy1 & $0.01\pm0.67$ & $41.84\pm0.88$ & $44.34\pm0.51$ & 42.47 & 10 \\
MCG-03-34-064 & 13:22:24.4 & $-$16:43:43 & 0.01654 & int-Sy &  $0.72\pm0.07$ & $45.06\pm0.15$ & $45.72\pm0.25$ & 43.18 & 1\\
MCG-03-58-007 & 22:49:36.9 & $-$19:16:24 & 0.03146 & int-Sy &  $0.48\pm0.07$ & $45.18\pm0.02$ & $45.07\pm0.05$ & 43.15 & 1\\
MCG-06-30-015 & 13:35:53.7 & $-$34:17:45 & 0.00775 & int-Sy &  $0.60\pm0.02$ & $43.84\pm0.01$ & $43.54\pm0.06$ & 42.8 & 4\\
Mrk509 & 20:44:09.7 & $-$10:43:25 & 0.03440 & int-Sy &  $0.83\pm0.02$ & $45.08\pm0.02$ & $45.00\pm0.05$ & 44.12 & 4\\
Mrk897 & 21:07:45.8 & 03:52:40 & 0.02634 & non-Sy &  $0.08\pm0.02$ & $44.39\pm0.01$ & $43.82\pm0.10$ & 42.0 & 1\\
Mrk1239 & 09:52:19.1 & $-$01:36:43 & 0.01993 & int-Sy&  $0.87\pm0.06$ & $44.89\pm0.03$ & $44.29\pm0.08$ & 43.32 & 4\\
NGC0034 & 00:11:06.5 & $-$12:06:26 & 0.01962 & Sy2&  $0.19\pm0.10$ &  $44.61\pm0.05$ & $<43.91$ & 41.69 & 1\\
NGC0424 & 01:11:27.5 & $-$38:05:01 & 0.01176 & int-Sy&  $0.80\pm0.03$ & $44.77\pm0.01$ & $44.42\pm0.03$ & 42.29 & 1\\
NGC0526A & 01:23:54.2 & $-$35:03:56& 0.01910 & int-Sy &  $0.40\pm0.11$ & $43.89\pm0.17$ & $44.48\pm0.16$ & 43.28 & 4\\
NGC1125 & 02:51:40.4 & $-$16:39:02 &  0.01093 & int-Sy &  $0.28\pm0.13$ & $43.84\pm0.08$ & $43.69\pm0.03$ & 41.94 & 1\\
NGC1194 & 03:03:49.2 & $-$01:06:12 & 0.01360 & Sy1 &  $0.79\pm0.04$ & $44.74\pm0.04$ & $43.88\pm0.15$ & 42.47 & 4\\
NGC1320 & 03:24:48.7 & $-$03:02:33 & 0.0089 & Sy2 & $0.56\pm0.01$ & $44.16\pm0.06$ & $43.95\pm0.02$ & 41.91 & 1\\
NGC1365 & 03:33:36.4 & $-$36:08:25 & 0.00546 & int-Sy &  $0.12\pm0.60$ & $44.05\pm0.52$ & $43.48\pm0.11$ & 42.12 & 4, 5\\
NGC1566 & 04:20:00.6 & $-$54:56:17 & 0.00500 & Sy1 &  $0.05\pm0.33$ & $42.81\pm0.16$ & $41.29\pm0.32$ & 40.97 & 4, 6\\
NGC2992 & 09:45:42.0 & $-$14:19:35 & 0.0077 & int-Sy & $0.35\pm0.03$ & $43.71\pm0.06$ & $44.15\pm0.11$ & 43.0 & 1\\
NGC4593 & 12:39:39.4 & $-$05:20:39 & 0.00900 & Sy1 &  $0.47\pm0.08$ & $43.16\pm0.01$ & $43.45\pm0.04$ &42.86 & 4, 7\\
NGC4602  & 12:40:36.5 & $-$05:07:55 & 0.0085 & int-Sy & $0.12\pm0.31$ & $43.70\pm0.11$ & $42.55\pm0.24$ & 42.06 & 1\\
NGC5135 & 13:25:44.0 & $-$29:50:02 & 0.01369 & Sy2 &  $0.25\pm0.04$ & $44.30\pm0.08$ & $44.09\pm0.02$ & 41.97 & 1\\
NGC5506 & 14:13:14.8 & $-$03:12:27 & 0.00618 & Sy2 &  $0.65\pm0.07$ & $44.21\pm0.07$ & $43.96\pm0.03$ & 42.74 & 1\\
NGC5995 & 15:48:24.9 & $-$13:45:28 & 0.02520 & int-Sy & $0.34\pm0.05$ & $44.90\pm0.02$ & $44.83\pm0.05$ & 43.37 & 1\\
NGC6810 & 19:43:34.1 & $-$58:39:21 & 0.00677 & non-Sy &  $0.13\pm0.52$ & $43.60\pm0.25$ & $<42.53$ &  39.6 & 11\\
NGC6860 & 20:08:46.1 & $-$61:05:56 & 0.01488 &Sy1 &  $0.49\pm0.05$ & $43.84\pm0.06$ & $43.80\pm0.01$ & 42.94 & 4\\
NGC6890 & 20:18:18.1 & $-$44:48:23 & 0.00810 & Sy2 & $0.13\pm0.69$ & $43.31\pm0.57$ & $43.45\pm0.01$ & 42.0 & 1\\
NGC7130 & 21:48:19.5 & $-$34:57:09 & 0.01615 & Sy2&  ... & ... & $44.47\pm0.01$ & $43.2$ & 1\\
NGC7213 & 22:09:16.2 & $-$47:10:00 & 0.00580 & Sy1 &  $0.44\pm0.01$ & $43.30\pm0.02$ & $<42.50$ & $42.1$ & 2\\
NGC7469 & 23:03:15.6 & 08:52:26 & 0.01632 & Sy1 &  $0.05\pm0.60$ & $44.51\pm0.41$ & $44.82\pm0.12$ & $43.5$ & 8\\
NGC7496 & 23:09:47.2 & $-$43:25:40 & 0.00550& non-Sy&  $...$ & $...$ & $<43.80$ &  ... & ... \\
NGC7603 & 23:18:56.6 & 00:14:38 & 0.02952 & Sy1 &  $0.55\pm0.06$ & $45.08\pm0.02$ & $,43.80$ & $43.56$ & 9\\
NGC7674 & 23:27:56.7 & 08:46:45 & 0.02892 & int-Sy &  $0.58\pm0.60$ & $45.31\pm0.30$ & $45.62\pm0.17$ & $44.02$ &  1\\
TOLOLO1238-364 & 12:40:52.9 & -36:45:22 & 0.01092 & int-Sy&  $0.32\pm0.60$ & $44.18\pm0.30$ & $44.31\pm0.06$ & $43.4$ & 1\\
\hline                                  
\end{tabular}
\tablefoot{ {\bf References:} 1. Salvestrini et al., in prep.; 2. \cite{Salvestrini2020}; 3. \cite{Ghosh2020}; 4. \cite{Asmus2015}; 5. \cite{Rivers2015}; 6.  \cite{Kawamuro2013}; 7. \cite{Middei2019}; 8. \cite{Mehdipour2018}; 9. \cite{Singh2011}; 10. \cite{Liu2014}; 11. \cite{Strickland2007}

}
\end{sidewaystable*}
%\end{table*}

\subsection{SALT RSS optical spectra}\label{sec:salt}

The optical spectra were obtained with the \textit{Robert Stobie} Spectrograph \citep[RSS;][]{Burgh2003,Kobulnicky2003} on SALT between May and November 2018 (program 2018-1-SCI-029, PI: L. Marchetti) and between November and June 2021 (program 2020-2-MLT-006, PI: L. Marchetti). 
Due to the weather conditions, the spectra of one of our targets (MCG+00-29-023) had a low signal-to-noise ratio, making impossible the use for the purposes of our science. Therefore, we excluded this source from our sample, focusing the analysis on 42 Seyfert galaxies.

The spectra were obtained using the RSS long-slit  mode, with a 1.5\arcsec slit and the PG0900 grating. This setting provides a spectral resolution of 5 \AA\ at 5000 \AA. The spectral range of our observational set-up was 4900-7900 \AA. We took two exposures (100 s each) per target, dithered along the slit by about 20\arcsec. 
In addition to the data set studied in this work, the program 2020-2-MLT-006 was designed to obtain for each target RSS spectra covering the bluer spectral range, down to 3600 \AA. The analysis of these new data will be the subject of future work. 
 
Initial data reduction steps (gain correction, cross-talk correction, overscan bias subtraction, and amplifier mosaicking) were performed by the SALT Observatory staff using a pipeline from the \texttt{PySALT} tool\footnote{\url{https://astronomers.salt.ac.za/software/pysalt-documentation/}} \citep{Crawford2010, Crawford2017}. The remaining steps were carried out in a standard way using \texttt{IRAF} tasks (arc line identification, wavelength calibration, background subtraction). Two consecutive exposures were combined to increase the signal-to-noise ratio and to remove the cosmic ray effects using the \texttt{L.A.Cosmic} task\footnote{\url{http://www.astro.yale.edu/dokkum/lacosmic/}} in \texttt{IRAF} \citep{vanDokkum2001}. Flux calibration was performed using spectrophotometric standard stars observed during twilight time. This refers to relative flux calibration only, since the absolute flux calibration is not feasible with SALT data alone as the unfilled entrance pupil of the telescope moves during the observations. Possible caveats related to the analysis of optical spectra and line flux calibration (see also Sect. \ref{sec:oiii_fluxcalibration}) are addressed in Sect. \ref{sec:oiii_caveats}.

\subsection{Mid-IR spectroscopy}\label{sec:ir}

The {\it Spitzer}/IRS high-resolution spectra ($10-37$ \mum) for our targets and their mid-IR line measurements are part of a larger collection of local AGN spectra by \cite{Tommasin2008, Tommasin2010}. The authors performed a careful background subtraction and accounted for the source extent using the ratio between the fluxes at 19\mum\ from the larger aperture of the long-high (LH) module of the IRS spectrograph and that of the short-high (SH) module. 
Additional analysis of these mid-IR emission features, combined with far-IR spectroscopic information (from {\it Herschel}/PACS and \textit{Herschel}/SPIRE spectrometers) are presented in \cite{Spinoglio2015} and \cite{Fernandez-Ontiveros2016}. We refer to \cite{Tommasin2008, Tommasin2010} and \cite{Fernandez-Ontiveros2016} for details on the determination of the fluxes used in this work, also reported in Table 1 of G16.  

 \cite{Tommasin2008, Tommasin2010} proposed a method for inferring the AGN bolometric luminosities from the IR line of \nevl, using the relation between the  \nevl\ line luminosity and the monochromatic luminosity at 19 \mum,\ which is, in turn, is converted into bolometric power from accretion following the relation by \cite{Spinoglio1995}. This can be performed in an analogous way using the \oivl\ line. We report in Table \ref{table:prop} the bolometric luminosities inferred from \oivl\ and \nevl\ for our targets taken from Table 3 of G16. These quantities are used in Sect. \ref{sec:opt_MIR}.
 
\subsection{X-ray data}\label{sec:X}

X-ray data from \emph{NuSTAR}, \textit{XMM}-\emph{Newton}, or \emph{Chandra} are available for 40 galaxies of our sample. 
We used X-ray 2-10 keV intrinsic luminosity (i.e., corrected for absorption) obtained from proprietary data or new analysis of published data (Salvestrini et al., in prep.) and, when not comprised in this analysis, from the literature \citep[][see Table \ref{table:prop} for details]{Ghosh2020, Asmus2015, Rivers2015, Kawamuro2013, Middei2019, Mehdipour2018, Singh2011, Liu2014, Strickland2007, Salvestrini2020}.

We preferentially used X-ray fluxes computed including \emph{NuSTAR} data (available for 37 Seyferts) as the \emph{NuSTAR} observations have proved fundamental to probe the primary X-ray emission in local AGN, even in the case of heavy obscuration ($N_{H}>10^{24}$ cm$^{-2}$), and to place constraints on the Compton thick AGN fraction \citep[e.g.,][]{Vignali2018, Marchesi2018, Marchesi2019, Torres-Alba2021}. This is because of the broader energy range (nominally, 3-79 keV) of \emph{NuSTAR}, compared to other X-ray instruments sensitive at energies below 10 keV, like \textit{XMM}-\emph{Newton} or \emph{Chandra}. 

No significant  X-ray variability has been observed in the Seyferts of our sample, with the exception of a couple of objects (NGC\,1365 and NGC\,2992), which showed evidence for relatively short-timescale variability, both in flux and spectral features \citep[see][and Salvestrini et al., in prep. for further details]{Rivers2015}.
To mitigate the effect of X-ray variability (which accounts for up to a factor of a few in terms of intrinsic luminosity for both objects) on the results, we consider the mean value of the 2-10 keV luminosity reported in the literature.
New analysis of \emph{NuSTAR} data for 25 of our targets will be presented in Salvestrini et al., in prep., the luminosity of the remaining sources are taken from the literature, as reported in Table \ref{table:prop}.

%--------------------------------------------------------------------
\section{Analysis of optical spectra}\label{sec:spectral_analysis}

\subsection{1D spectral extraction}

Starting from the reduced 2D SALT data, we extracted 1D spectra using a slit length of 11$\arcsec$ to match that of the {\it Spitzer}/IRS SH mode. We applied the correction for interstellar galactic extinction using the maps from \cite{Schlegel1998}, with the updated calibration from \cite{Schlafly&Finkbeiner2011}, and the \cite{Fitzpatrick1999} extinction function assuming an R$_{\rm V}$ of 3.1.

\subsection{Emission line measurements}\label{sec:line_measurements}
We measured the emission lines after subtracting the stellar continuum with the \textsc{PYTHON} version of the {\tt ppxf} code \citep{Cappellari2017}.  To model the continuum we adopted the stellar libraries from \cite{Vazdekis2010}, downloaded from the MILES website \footnote{\url{http://miles.iac.es/pages/stellar-libraries/miles-library.php}}. These templates are provided in the spectral wavelength range $3525-7500$ \AA\ and sampled at a spectral resolution of a FWHM=2.5 \AA.

We computed emission line intensities with the {\tt pyspeclines} package \footnote{\url{https://github.com/jacopo-chevallard/PySpecLines}}, which makes use of the \textsc{PYTHON} package \textsc{pyspekit} \citep{Ginsburg2011} and fits a Gaussian (or multiple Gaussians) using a Markov chain Monte Carlo method to compute the errors.  
When broad hydrogen lines were present in the spectra, as in the case of Seyfert 1, or when multiple kinematics components are required to reproduce a line, as in the case of \oiiirl, we used multiple Gaussian components with different widths. We required the width of the narrower component to be the same as that of the other narrow forbidden lines. 
Regarding the \oiiirl\ line, 14/42 targets required a fit with two components, instead of a single one, to match the emission line profile. In these cases, we do not attribute a physical meaning to these components as the observed emission lines are dominated by AGN photoionization (see Fig. \ref{fig:BPT}). We did not find strong signs for the line profiles to be dominated by kinematically disturbed outflowing gas, like offset from the systemic velocity or double-peaked profiles \citep[e.g.,][]{Davies2020}. However, in Sect. \ref{sec:shock} we discuss the impact of potential contribution from shocks, due, for example, to outflows of galactic winds, to the line emitted spectra. 

We measured all the main optical lines, namely \hbl, \oiiibl, \oiiirl, \oibl, \oirl, \niibl, \hal, \niirl, \siibl,\ and \siirl.
Hereafter, we refer to \hbl, \hal, \oiiirl, \oirl, \niirl, and  \siibl$+$ \siirl\ as \hb, \ha, \oiii, \oi, \nii, and \sii, respectively, unless stated otherwise. 
We corrected the line fluxes for attenuation by dust using the Balmer decrement from the \ha/\hb\ ratio when both lines are available, assuming a \cite{Cardelli1989} reddening curve and a case B recombination ratio of 3.1, characteristic of the physical conditions of the narrow-line-emitting regions (NLRs) of AGN \citep{Kewley2006, Groves2012, PerezDiaz2022}. A detailed discussion on the \ha/\hb\ ratio in AGN can be found in \cite{Armah2021}. This is possible only for 25 of our targets as the remaining 17 have \hb\ in absorption or outside the spectral coverage. When no \ha/\hb\ ratio is available, we used the attenuation from the fitting to the broadband photometry presented in G16 and obtained assuming a two-component model of dust attenuation by \cite{Charlot2000}.
 We point out that there is a substantial scatter ($\approx 2$ mag) between the two dust attenuation measurements (when both are available) but with the lack of the \ha/\hb\ information for the whole sample we had to rely on the results from the SED fitting (see Sect. \ref{sec:dust_caveats} for a more detailed discussion). This choice could partially affect the results in Sects. \ref{sec:opt_Xray} and \ref{sec:opt_MIR} but impact only marginally the remainder of the work as we consider only ratios between lines very close in wavelengths (see Sect. \ref{sec:dust_caveats} for a more quantitative discussion). Table \ref{table:opt_ratios} lists the main dust attenuation corrected line ratios that will be used later in Sect. \ref{sec:bpt} and the V-band attenuation, A$_{\rm v}$, applied to the line fluxes and ratios. It should be noted that in the case of Seyfert 1 and intermediate Seyfert we considered only the narrow component of the Balmer lines to compute the line ratios. 

\subsection{\oiiirl\ flux calibration}\label{sec:oiii_fluxcalibration}

While a relative (shape conservative) flux calibration has been performed during the data reduction (Sect. \ref{sec:salt}), an absolute flux calibration is not possible with SALT data alone due to the variable pupil of the telescope. This prevents us from using line fluxes (e.g., \oiiirl). To obtain an absolute flux calibration we considered aperture photometry in the V-band from \cite{Hunt1999b}, available for 28 of our 42 targets.   
We measured the V-band magnitude in the 1D SALT spectra (prior stellar continuum subtraction) and compared it with the V-band aperture photometry from \cite{Hunt1999b}, obtaining a multiplicative factor, f$_{\rm cor}$, to be applied to the line fluxes measured on the SALT spectra. In particular, we applied this to the \oiii\ line to obtain an absolute \oiii\ flux. To the remaining 14 targets with no aperture photometry, we applied an average relation obtained through an orthogonal distance regression (ODR) fitting to the magnitude computed in the SALT spectra and that from \cite{Hunt1999b}. We report the multiplicative factors, f$_{\rm cor}$, and the dust attenuation corrected absolute \oiii\ luminosities (i.e., after being corrected by dust attenuation and applying the multiplicative factor) in Table  \ref{table:opt_ratios}. 

As a consistency check, we compared our measurements of the \oiii\ fluxes with those from \cite{Malkan2017}, collected from optical spectra available in the literature for 185 Seyfert galaxies, including nearly all those from the 12MGS. The \oiii\ flux measurements from \cite{Malkan2017} are available for 40/42 objects of our sample. Even though the spectra from  \cite{Malkan2017} are heterogenous and use different spectral apertures, the \oiiirl\ fluxes reported in their Table 4 are consistent with those computed from the SALT spectra, with 70\% of the targets within $1\sigma$ from the 1:1 relation. We further discuss the implications of our method for obtaining an absolute line flux in Sect. \ref{sec:oiii_caveats}. 

%\begin{sidewaystable*}
\begin{table*}
\caption{Emission line fluxes of the main optical lines}        
\label{table:opt_ratios}     
\centering                          
\begin{tabular}{l c c c c c c c}       
\hline\hline                 
Name &  [\ion{O}{iii}]/ H$\beta$ & [\ion{O}{iii}]/H$\alpha$ & [\ion{N}{ii}]/H$\alpha$  & [\ion{S}{ii}]/H$\alpha$ & A$_{v}$ & f$_{\rm cor}$([\ion{O}{iii}]) & log([\ion{O}{iii}])   \\    
& & & & & & log(erg/s)  \\
\hline                        
CGCG381-051  & $0.235\pm0.002$ & $0.027\pm0.001$ & $0.540\pm0.001$ & $0.212\pm0.001$ & 2.7 & 3.80 & 39.78\\    
ESO033-G002  & $5.358\pm0.033$ & $0.128\pm0.001$ & $1.274\pm0.003$ & $0.714\pm0.002$ & 3.7 & 1.67 & 41.22\\
ESO141-G055 & ... & ... &  ... & ... & 1.2 &  1.02 & 41.49\\
ESO362-G018  & $4.470\pm0.005$ & $0.100\pm0.000$ & $0.353\pm0.001$ & $0.204\pm0.001$ & 0.4 & 1.31 & 41.01 \\
IC4329A & $11.100\pm0.104$ & $0.099\pm0.001$ & $0.435\pm0.005$ & $0.236\pm0.012$ & 4.2 & 3.40& 41.20\\
IC5063 & ... & $0.100\pm0.001$ & $0.649\pm0.005$ & $0.510\pm0.006$ & 0.9 & 4.11 & 41.33\\
IRASF01475-0740 & $5.382\pm0.047$ & $0.118\pm0.001$ & $0.608\pm0.001$ & $0.166\pm0.002$ & 2.4 & 3.12& 40.59\\
IRASF03450+0055 & $0.330\pm0.010$ & $...$ & $...$ & $...$ & 1.0 & 1.39& 40.69\\
IRASF04385-0828 & $2.868\pm0.033$ & $0.161\pm0.001$ & $0.871\pm0.003$ & $0.614\pm0.002$ &3.3 & 0.64 & 40.11\\
IRASF05189-2524 & $8.281\pm0.327$ & $0.146\pm0.015$ & $1.419\pm0.168$ & ... & 0.8 & 0.67 & 40.96\\ 
IRASF15480-0344 & $10.293\pm0.013$ & $0.114\pm0.001$ & $0.715\pm0.001$ & $0.281\pm0.001$ & 0.4 & 1.58 & 42.10\\
MCG-02-33-034 & $11.517\pm0.026$ & $0.120\pm0.001$ & $0.271\pm0.001$ & $0.666\pm0.001$ & 0.5 & 1.84 &41.14\\
MCG-03-34-064 & $14.415\pm0.181$ & $0.224\pm0.006$ & $2.917\pm0.083$ & $1.453\pm0.056$ & 2.4 & 1.31 & 41.98\\
MCG-03-58-007& $7.781\pm0.281$ & $0.096\pm0.001$ & $1.103 \pm0.002$ & $0.220\pm0.001$ &  0.4 & 1.35& 41.46\\
MCG-06-30-015 & ... & $0.052\pm0.001$ & $0.120\pm0.001$ &$0.296\pm0.001$ & 2.7 &1.54 & 40.09\\
Mrk509 & $7.119\pm0.012$ & $0.046\pm0.001$ & $0.476\pm0.003$ & ... & 1.6 & 2.21 & 41.70\\
Mrk897 & $0.470\pm0.009$ & $0.023\pm0.001$ & $0.460\pm0.004$ & $0.265\pm0.002$ & 0.9 & 2.13 & 40.33\\
Mrk1239 & $5.358\pm0.039$ & $0.012\pm0.003$ & $0.693\pm0.233$ & $0.102\pm0.003$ & 1.6 & 0.98 & 41.30\\
NGC0034 & $1.605\pm0.016$ & $0.163\pm0.001$ & $1.218\pm0.003$ & $0.748\pm0.012$ & 2.8 & 2.45 & 40.11\\
NGC0424 &... & $0.084\pm0.023$ & $0.617\pm0.170$ & $0.246\pm0.067$ & 0.8 & 1.59 & 40.92\\
NGC0526A & $9.584\pm0.038$ & $0.222\pm0.069$ & $1.269\pm0.405$ & $0.985\pm0.305$ & 1.2 & 2.38 & 41.37\\
NGC1125 & ... &  $0.135\pm0.003$ & $0.776\pm0.011$ & $0.714\pm0.008$ & 1.6 & 11.19 & 40.63\\
NGC1194 & ... & $0.0724\pm0.001$ & $0.678\pm0.001$ & $0.457\pm0.001$ & 3.2 & 3.75 & 40.90\\
NGC1320 & $33.552\pm0.770$ & ... & $0.639\pm0.020$ & $0.743\pm0.025$ & 1.4 &  2.92&40.43\\
NGC1365 & ... & $0.022\pm0.001$ & $0.532\pm0.002$ & $0.184\pm0.001$ & 1.6 & 2.27& 39.43\\
NGC1566 & ... & $0.175\pm0.021$ & $1.072\pm0.119$ & $0.486\pm0.048$ & 2.5 & 0.97& 39.63\\
NGC2992 & ... & $0.300\pm0.003$ & $0.905\pm0.001$ & $0.798\pm0.001$ & 0.9 & 2.17 & 40.67\\
NGC4593 & ... & $0.098\pm0.001$ & $1.063\pm0.003$ & $0.509\pm0.002$ & 1.8 &  1.39 &40.11\\
NGC4602 & ... & $0.042\pm0.001$ & $0.568\pm0.002$ & $0.327\pm0.001$ & 0.5 & 2.49 &38.64\\
NGC5135 & $3.928\pm0.066$ & $0.044\pm0.001$ & $1.286\pm0.002$ & $0.698\pm0.001$ & 0.1 & 1.32 &41.00\\
NGC5506 & ... & $0.152\pm0.001$ & $0.983\pm0.001$ & $0.771\pm0.001$ & 1.2 & 6.69 & 40.74\\
NGC5995 & $2.812\pm0.033$ & ... & $0.862\pm0.006$ & ... & 2.3 & 3.30 & 40.67\\
NGC6810 & ... & $0.022\pm0.000$ & $0.588\pm0.000$ & $0.225\pm0.000$ & 1.2 & 0.84 & 39.26\\
NGC6860 & $2.554\pm0.005$ & $0.113\pm0.001$ & $0.756\pm0.003$ & $0.466\pm0.002$ & 1.4 & 2.79 & 40.70\\
NGC6890 & ... & $0.060\pm0.001$ & $0.691\pm0.001$ & $0.265\pm0.001$ & 1.5 & 1.69 & 40.13\\
NGC7130 & $3.590\pm0.001$ & $0.065\pm0.00$ & $0.938\pm0.001$ & $0.268\pm0.001$ & 1.3 & 2.40 & 41.16\\
NGC7213 & ... & $1.413\pm0.005$ & $1.910\pm0.007$ & $2.029\pm0.007$ & 1.3 & 1.29 & 39.79\\
NGC7469 & $4.313\pm0.022$ & $0.049\pm0.000$ & $0.481\pm0.001$ & $0.243\pm0.001$ & 1.4 & 1.79 & 41.08\\
NGC7496 & ... & $0.042\pm0.001$ & $0.531\pm0.003$ & $0.298\pm0.001$ & 0.9 & 1.34 & 39.53\\
NGC7603 & $1.044\pm0.001$ & $0.085\pm0.002$ & $0.473\pm0.010$ & $0.305\pm0.007$ & 0.4 & 1.65 & 40.78\\
NGC7674 & $15.954\pm0.028 $ & $0.043\pm0.001$ & $0.858\pm0.001$ & $0.480\pm0.001$ & 2.0 & 2.18 & 41.94\\
TOLOLO1238-364 & ... & $0.132\pm0.001$ & $0.643\pm0.001$ & $0.490\pm0.001$ & 4.3 & 1.08 & 40.76\\

\hline                                  
\end{tabular}
%\end{sidewaystable*}
\end{table*}

\section{\oiiirl\ and other tracers of AGN activity}\label{sec:lacc_tracers}

We first investigate how our targets populate optical line-ratio diagrams commonly used to identify AGN (Sect. \ref{sec:bpt}) and then investigate the correlations between the attenuation-corrected \oiiirl\ luminosity and other tracers of AGN activity. The \oiiirl\ line is one of the most prominent optical line in AGN, less contaminated by star formation and, in turn, commonly used as isotropic indicator of the intrinsic AGN strength \citep[e.g.,][]{Bassani1999, Heckman2005, Lamassa2010}. In particular, we first verify how our sample compares with the already known relation between  \oiiirl\ and X-ray luminosities (Sect. \ref{sec:opt_Xray}) and then explore how the \oiiirl\ luminosity compares with that of high-ionization IR emission lines (Sect. \ref{sec:opt_MIR}). 

\subsection{Optical diagnostic diagrams}\label{sec:bpt}

\begin{figure*} 
\centering
  \includegraphics[width=17cm]{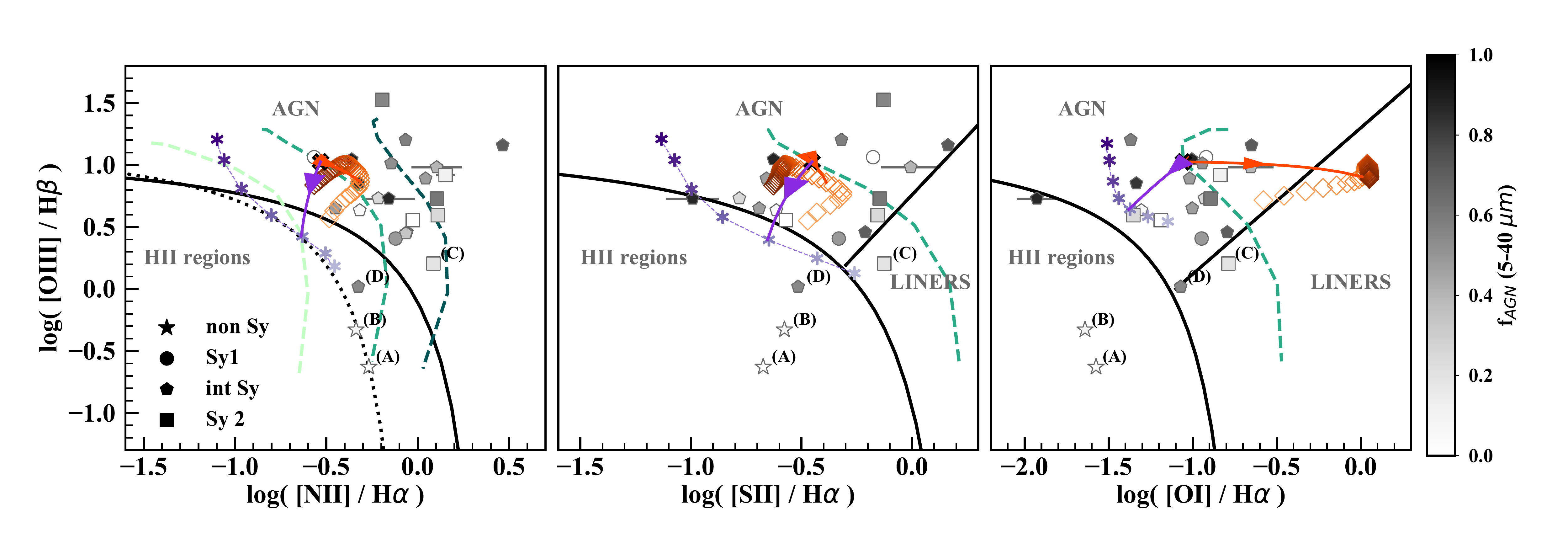}
     \caption{Optical line ratio diagrams, \oiii/\hb\  versus \nii/\ha, \sii/\ha, and \oi/\ha\ (from left to right). Observations are color-coded with gray shades based on the AGN contribution to the mid-IR ($5-40 \mu$m), f$_{\rm AGN}$. Different symbols indicate the classification of the targets (as labeled in the legend). The solid black curves are the criteria for separating AGN from \ion{H}{ii} regions and  LINERS proposed by \cite{Kewley2006}. The dotted curve is the demarcation line between AGN and \ion{H}{ii} regions from \cite{Kauffmann2003}. The dashed green curves indicate AGN models from F16 (Sect. \ref{sec:photomod})  with Z$=0.017$, $n_{\rm H}=10^3$cm$^{-3}$, $\xi_{\rm d}=0.3,$  $\alpha=-1.7,$ and the ionization parameter \logU varying from $-4.5$ to $-1.5$ (from bottom to top). In the left panel, dashed lighter and darker green curves are for AGN models of $Z=0.008$ and $Z=0.04$.
      The arrows illustrate the effect of adding to an AGN model with Z$=0.017$,  \logU$= -2.5$, $n_{\rm H}=10^3$cm$^{-3}$, $\xi_{\rm d}=0.3,$ and $\alpha=-1.7$ (black cross) a 0\% to 90\% fractional contribution from star formation (violet) and shocks (orange) to the total \hb\ line emission. The asterisks (empty diamonds) indicate predictions of AGN$+$SF (AGN$+$shocks) models with 90\% contribution to the total \hb\ line from the star formation (shocks). The ionization parameter of the stellar models increases from $log ( \langle U^{\star} \rangle ) = -3.0$ (from the lighter to the darkest violet shade). The shock velocity increases from 200 to 1000 km/s (from the lighter to the darkest orange shade).  }
       \label{fig:BPT}
\end{figure*}

Diagnostic diagrams based on ratios of strong optical emission lines, like \hb, \ha, \oiii, \oi, \nii,\ and \sii,  are routinely used to differentiate AGN activity from star formation.
In Fig. \ref{fig:BPT} we consider three commonly used diagnostic diagrams based on the \oiii/\hb, \nii/\ha, \sii/\ha, and \oi/\ha\ emission line ratios, originally presented in BPT (\citealt*{BPT}) and \cite{Veilleux1987}. We explore the position of our sample in these diagrams. One can note that only a fraction of the 42 targets are shown in Fig. \ref{fig:BPT} because not all the emission lines of interest were detected in the SALT spectra of all the objects (see Table \ref{table:opt_ratios}).

The two non-Sy targets that we could place in these diagrams (i.e., CGCG381-051 and Mrk\,897, flagged with (A) and (B), respectively, in Fig. \ref{fig:BPT}) lie in the \ion{H}{ii}-region locus below both the \cite{Kewley2006} and \cite{Kauffmann2003} demarcation curves (solid and dotted curves, respectively). These targets were originally classified as Seyfert galaxies but their classification was revised to non-Seyfert by \cite{Tommasin2010}. Our SALT data are in line with this latter classification. 
All of the Sy-type sources, with the exception of two targets (NGC\,0034 and NGC\,7603, flagged with (C) and (D), respectively, in Fig. \ref{fig:BPT}) lie in the AGN-dominated area in all three diagrams, according to the criteria from \cite{Kewley2006}.
NGC\,7603 is in the composite area between the theoretical demarcation line of maximum starburst from \citet[][solid curve in Fig. \ref{fig:BPT}]{Kewley2006} and the empirical AGN-\ion{H}{ii} separation criteria from \citet[][dotted curve]{Kauffmann2003} in the \oiii/\hb\ versus \sii/\ha\ and  \oiii/\hb\ versus \oi/\ha\ diagrams. In these same diagrams, NGC\,0034, whose X-ray spectrum shows evidence of a heavy obscured AGN (Salvestrini et al., in prep.), is just at the border between the AGN-dominated region and the area populated by LINERS and by objects where the ionization from shocks or stars in the post-asymptotic-giant-branch phase dominate the emission line spectra.

\subsection{\oiiirl\ and X-ray luminosities}\label{sec:opt_Xray}

The 2-10 keV X-ray luminosity is commonly used as proxy of the intrinsic luminosity of the AGN power and found to correlate with the \oiii\ luminosity over a wide range of magnitudes and for different AGN types \citep[e.g.,][]{Netzer2006, Panessa2006, Lamastra2009, Georgantopoulos2010}.

\begin{figure*}
\centering
   \includegraphics[width=17cm]{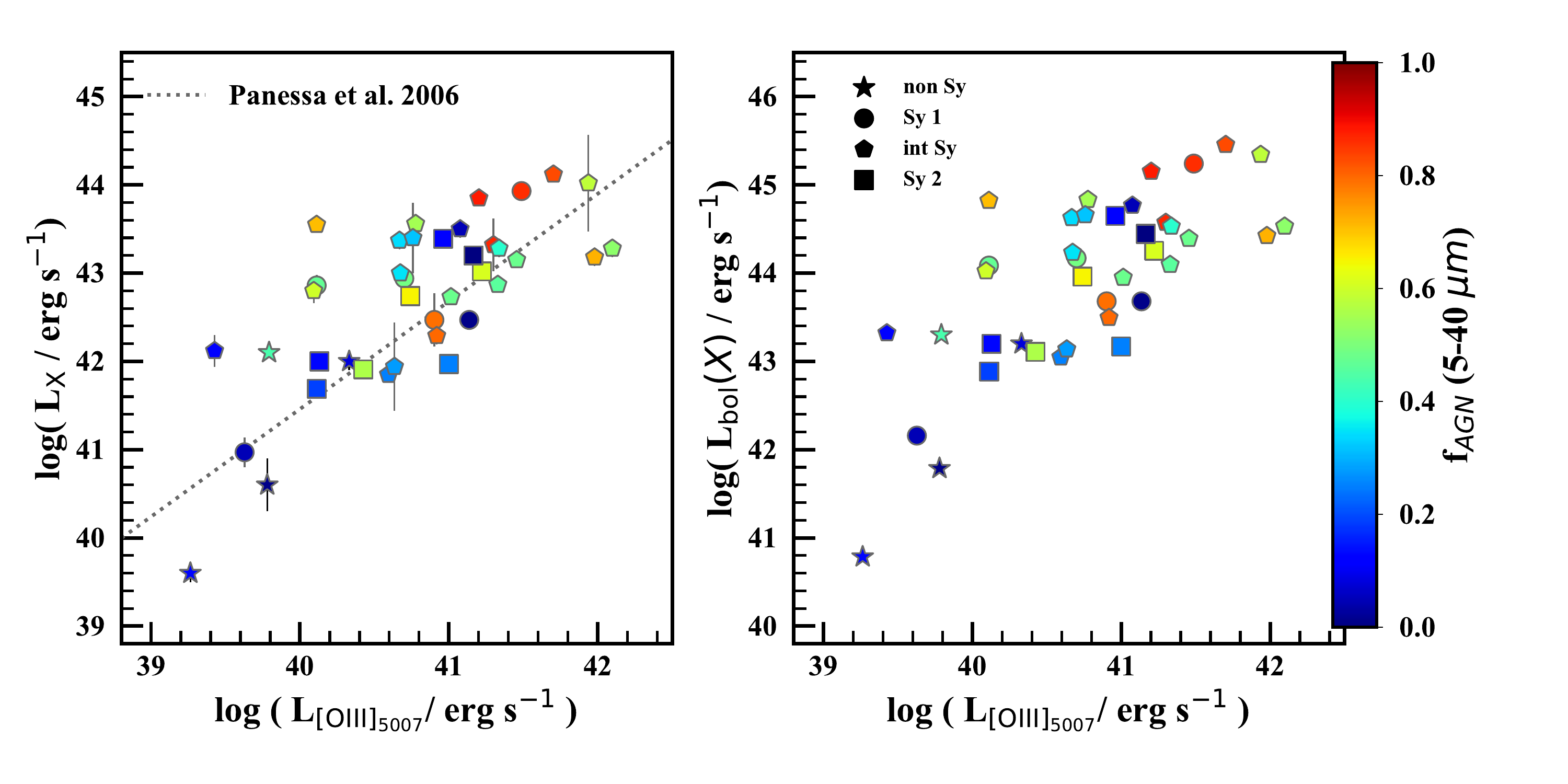}
     \caption{\oiii\ compared with X-ray (left) and bolometric (right) luminosities. Observed data are color-coded as a function of the AGN contribution to the mid-IR ($5-40 \mu$m), f$_{\rm AGN}$, inferred via SED fitting. Different symbols indicate the classification of the targets (as labeled in the legend in the right panel). The dotted line is the relation from \cite{Panessa2006}, while the bolometric luminosity, L$_{\rm bol}$, is obtained using the prescription from \cite{Duras2020}. }
     \label{fig:LxLoiii}
\end{figure*}

The left panel of Fig. \ref{fig:LxLoiii} shows the absorption-corrected X-ray luminosities versus the dust attenuation-corrected \oiii\ luminosities of our targets, distinguishing between different spectral types (as indicated by the different symbols). 
Our targets follow the relation from \cite{Panessa2006}, which was obtained for a heterogenous sample of local AGN as the sample discussed in the present work, over a wide range of 2-10 keV X-ray luminosities, from $\approx10^{38}$ to $\approx 10^{43}$ erg/s. Four targets in our sample are spectroscopically classified as non-Seyfert (Fig. \ref{fig:BPT}, star symbols), as detailed in \cite{Tommasin2008}, and the nebular emission from H{\sc ii} regions is expected to contribute significantly to their \oiii\ flux. We note that these targets agree well with the \cite{Panessa2006} relation, with NGC\,6810 showing the weakest fluxes. 

We inferred the AGN bolometric luminosity, ${\rm L_{bol}(X)}$, from the X-ray luminosity using the prescription for the bolometric corrections from \citet[][see also \citealp{Lusso2012}]{Duras2020}. The \oiiirl\ luminosity increases for higher X-ray bolometric luminosities, along with the AGN contribution to the mid-IR emission (right panel of Fig. \ref{fig:LxLoiii}).

\subsection{\oiiirl\ and high-ionization ($\gtrsim 40$ eV) mid-IR lines}\label{sec:opt_MIR}

\begin{figure}
\centering
   \includegraphics[width=9cm]{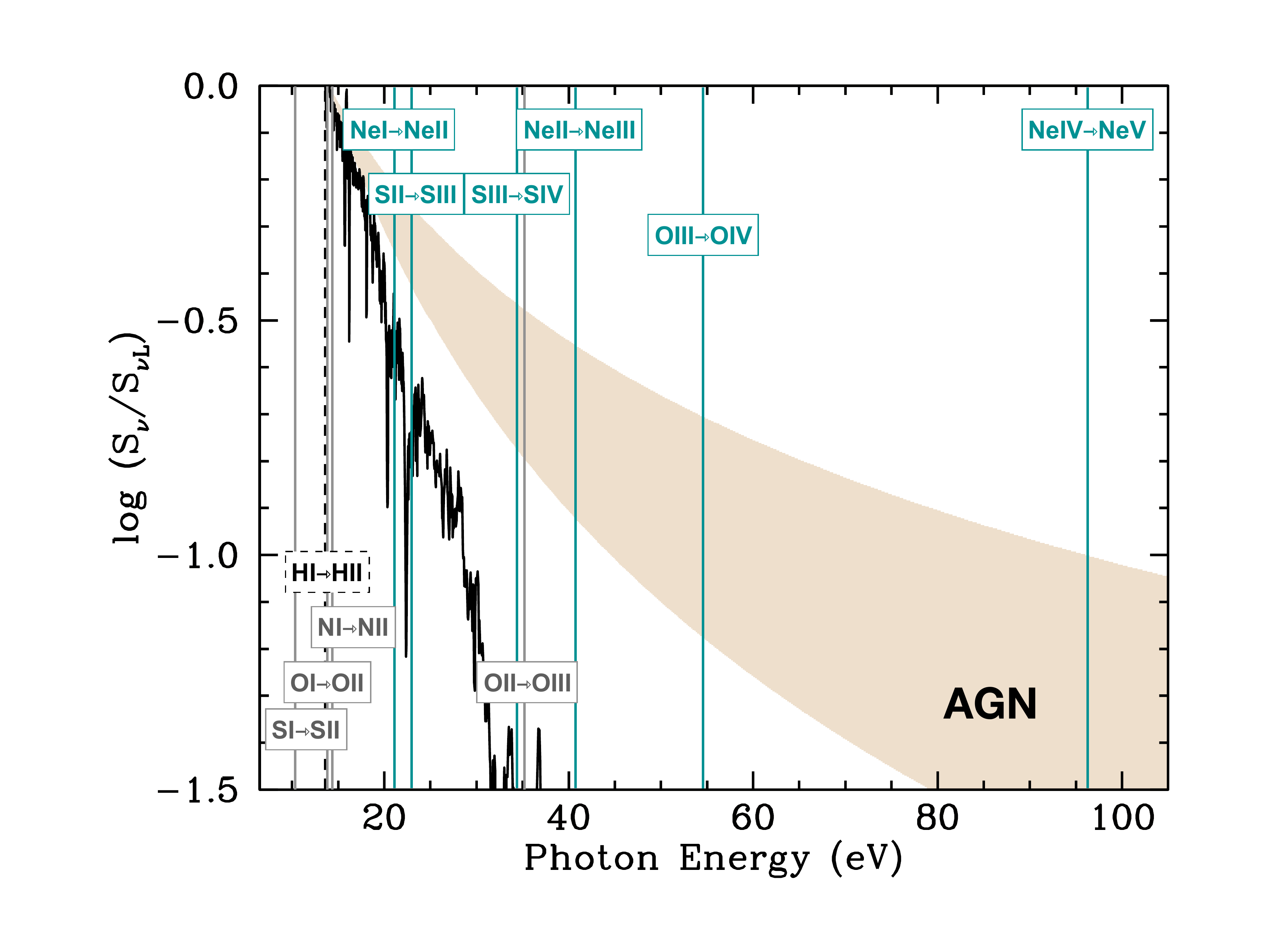}
     \caption{Examples of incident ionizing spectra (in units of the luminosity per unit frequency at the Lyman limit) as a function of photon energy in the AGN and the star-forming galaxy models described in Sect. \ref{sec:model_comparison}. The beige shaded area highlights the AGN accretion disk ionizing radiation with power-law indices between $\alpha = -2.0$ (bottom edge) and $-1.2$ (top edge). The black line shows the ionizing spectra of a star-forming galaxy with metallicity $Z=0.017$. Vertical lines indicate the ionizing energies of ions of different species (dashed line for hydrogen and continuous lines for the metal transitions considered in this work; gray and green for optical and mid-IR, respectively).}
     \label{fig:ionspec}
\end{figure}

High-ionization (above $\approx40$ eV) emission lines are good tracers of the AGN photoionization power because the ionizing radiation from stars does not produce a significant amount of hard enough ionizing photons to dominate the emission of these lines. We illustrate this in Fig. \ref{fig:ionspec}, where we compare the SEDs of the incident radiation (Sect. \ref{sec:photomod} for details and references) of an AGN and a stellar population of metallicity Z$=$0.017, close to the solar value (for reference the value of the present-day solar metallicity of the models is $Z_{\odot}=0.01524$). 

We considered mid-IR emission lines with ionization potentials higher than 40 eV, namely: \neiiil, \oivl,\ and \nevl\ (hereafter \neiii, \oiv,\ and \nev), with ionization potentials of 40.96, 54.9, and 97.1  eV, respectively). While ionization due to young massive stars can still partially contribute to the \neiii\ and \oiv\ lines, the \nev\ has been found to be a good tracer of AGN activity \citep{Abel2008}. Moreover, the energy of 97.1 eV required to quadruply ionize neon (Ne$^{4+}$ or \nev) is too high to be purely driven by star formation, except in some extreme cases where Wolf-Rayet stars are present \citep{Schaerer1999, Abel2008, Cleri2022}.

\begin{figure*}
\centering
 \includegraphics[width=9cm]{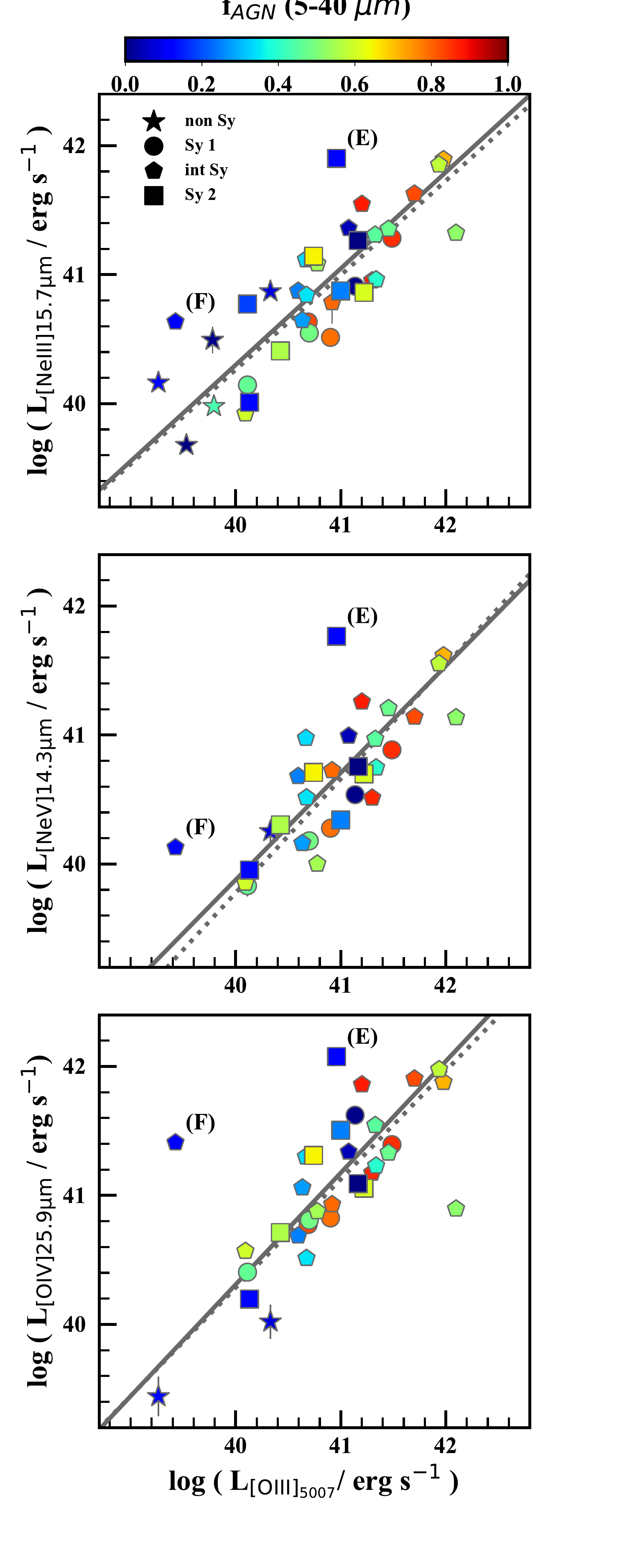}
  \includegraphics[width=9cm]{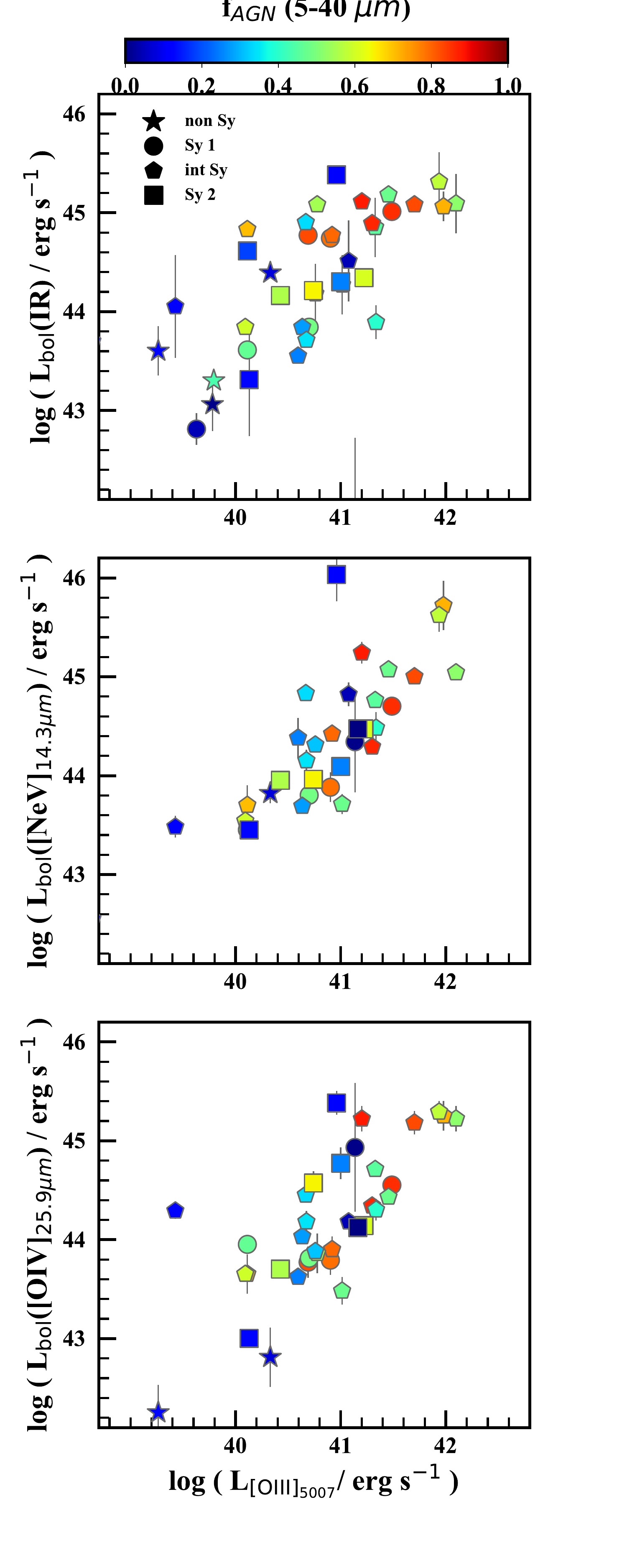}
     \caption{\oiii\ luminosities measured in the SALT spectra, after dust attenuation correction and flux calibration, compared with luminosities of mid-IR lines (\neiii, \nev, and \oiv, from top to bottom on the left) and with bolometric luminosities inferred in different ways (from SED fitting, \nev\ lines,\ and \oiv\ lines, from top to bottom on the right). Symbols and colors of the observations are the same as in Fig. \ref{fig:LxLoiii}. Solid lines show the relations in Eqs. \ref{eq:1}, \ref{eq:2}, and \ref{eq:3}, while dotted lines are those obtained considering the \oiii\ luminosities from \cite{Malkan2017}.}
     \label{fig:LoiiiLIR}
\end{figure*}

We compared the luminosities of these mid-IR lines with that of the optical \oiii\ emission line, which, even with a lower ionization potential ($35.11$ eV) than mid-IR transitions, is one of the strongest lines in the optical range and used as indicator of the AGN intrinsic power  \citep[e.g.,][]{Bassani1999, Heckman2005, Lamassa2010}. The left panels of Fig. \ref{fig:LoiiiLIR} show that the luminosities of \neiii, \nev\ and \oiv\  correlate with that of \oiii, with scatters of 0.57, 0.53, 0.54 dex, respectively. We performed ODR fitting and obtained the following relations:

\begin{equation}
\begin{aligned}
log_{10}( L_{[\ion{Ne}{iii}]} / erg \, s^{-1} ) = (0.768\pm0.074) \times  log_{10}( L_{[\ion{O}{iii}]} / erg\, s^{-1})\\
+ (9.530 \pm 3.003)
\end{aligned}\label{eq:1}
,\end{equation}

\begin{equation}
\begin{aligned}
log_{10}( L_{[\ion{Ne}{v}]} / erg \, s^{-1} ) = (0.770\pm0.089) \times  log_{10}( L_{[\ion{O}{iii}]} / erg \, s^{-1} )\\
+ (9.124 \pm 3.643)
\end{aligned}\label{eq:2}
,\end{equation}

\begin{equation}
\begin{aligned}
 log_{10}( L_{[\ion{O}{iv}]} / erg \, s^{-1} ) = (0.865\pm 0.134) \times log_{10}( L_{[\ion{O}{iii}]} / erg\, s^{-1}) \\
+ (5.670 \pm 5.465)
\end{aligned}\label{eq:3}
.\end{equation}

These relations could be useful for the design of IR observations of AGN, like those from JWST/MIRI and from future IR missions. 
Given that our sample is very heterogenous in terms of types of Seyfert, the relations reported above are proposed as average relations to be applied to local AGN. One would need a larger sample to obtain relations specific for a given class of objects.
The relations in Eqs. \ref{eq:1}, \ref{eq:2} and \ref{eq:3} are consistent with those obtained using the \oiii\ luminosities from the literature \citep[collected by][]{Malkan2017} instead of the \oiii\ luminosities measured in the SALT spectra, as shown by the dotted lines in Fig. \ref{fig:LoiiiLIR}.

We note that the fractional contribution from the AGN to the $5-40$ \mum\ mid-IR luminosity, f$_{\rm AGN}$, increases with the luminosity of the line that traces the intrinsic AGN power (Fig. \ref{fig:LoiiiLIR}). 
Two targets, namely IRASF05189-2524 (Sy2) and NGC\,1365 (int-Sy) flagged as (E) and (F), respectively, in Fig. \ref{fig:LoiiiLIR}, have \nev\ and \oiv\ luminosities higher than derived from the mean relations with \oiii.
IRASF05189-2524 is a major merger with potentially high contribution from shocks, in addition to the AGN, to the gas heating. This has been already observed studying the emission of the molecular gas of this target \citep{Pereira-Santaella2014}. Spatially resolved \oiii\ emission from MUSE/VLT observations of NGC\,1365 has revealed a kiloparsec-scale biconical outflow ionized by the AGN \citep{Venturi2018}. Shocks and outflows in systems as such could influence the relative luminosities of the lines, resulting in a higher deviation from the mean relations presented above. 
We explore the impact of fast radiative shocks on the optical and mid-IR line ratios in Sect. \ref{sec:shock}.

In the right panels of Fig. \ref{fig:LoiiiLIR}, the \oiii\ luminosity is compared with the AGN bolometric luminosities, $L_{\rm bol}$, computed in different ways, namely from the 1-1000 \mum\ rest-frame IR luminosity inferred by means of SED decomposition and from the \nev\ and \oiv\ high-ionization emission lines (see Sect. \ref{sec:ir} of this work and Sect. 4.2 of G16 for more details).
Even if the relation between the \oiii\ and the bolometric luminosity obtained via SED fitting is more dispersed than the relations involving mid-IR transitions,  probably due to the larger uncertainties related to the models and assumptions of the fitting itself, the objects with higher fractional contribution from the AGN to the $5-40$ \mum\ mid-IR luminosity have, on average,  higher bolometric luminosity (irrespective of the method used to compute it).

%--------------------------------------------------------------------

\section{Mid-IR line ratios and diagnostic diagrams}\label{sec:mir}

The  {\it Spitzer}/IRS spectra of our targets offer access to different ionization states of the gas through the detection of several forbidden emission lines (see Fig. \ref{fig:ionspec}), such as \neiil, \siiil, and \sivl\  (hereafter \neii, \siii, and \siv) in addition to the \neiii, \oiv,\ and \nev\ lines already explored in the previous section. These mid-IR transitions, being insensitive to heavy dust attenuation, are optimal tracers of the ionizing radiation. 

Several mid-IR line ratios have been proposed as diagnostics of star formation and AGN activity, like the \nev/\neii, \oiv/\neii,\ and \oiv/(\neiii+\neii) single ratios \citep[e.g., ][]{Genzel1998, Fernandez-Ontiveros2016} and the \nev/\neiii\ versus \neiii/\neii\ diagnostic diagram \citep{Groves2006b}.
Other diagnostic diagrams useful to explore the contribution of star formation are \siv/\siii\ and \siv/\neii\ versus \neiii/\neii\ \citep{Inami2013, Cormier2015}.

Another mechanism able to ionize the ISM of galaxies is radiative shocks. \cite{Allen2008} present a study on how models of fast radiative shocks populate a set of mid-IR diagnostics diagrams. 

In this section we discuss the application of mid-IR diagnostics based on several line ratios to our sample of Seyfert galaxies.
We then compare the observed line ratios with predictions of the line emission from the NLR photoionized gas in AGN and consider the additional contributions from stars and fast radiative shocks in Sect. \ref{sec:model_comparison}. 

\begin{figure*}
\centering
  \includegraphics[width=17cm]{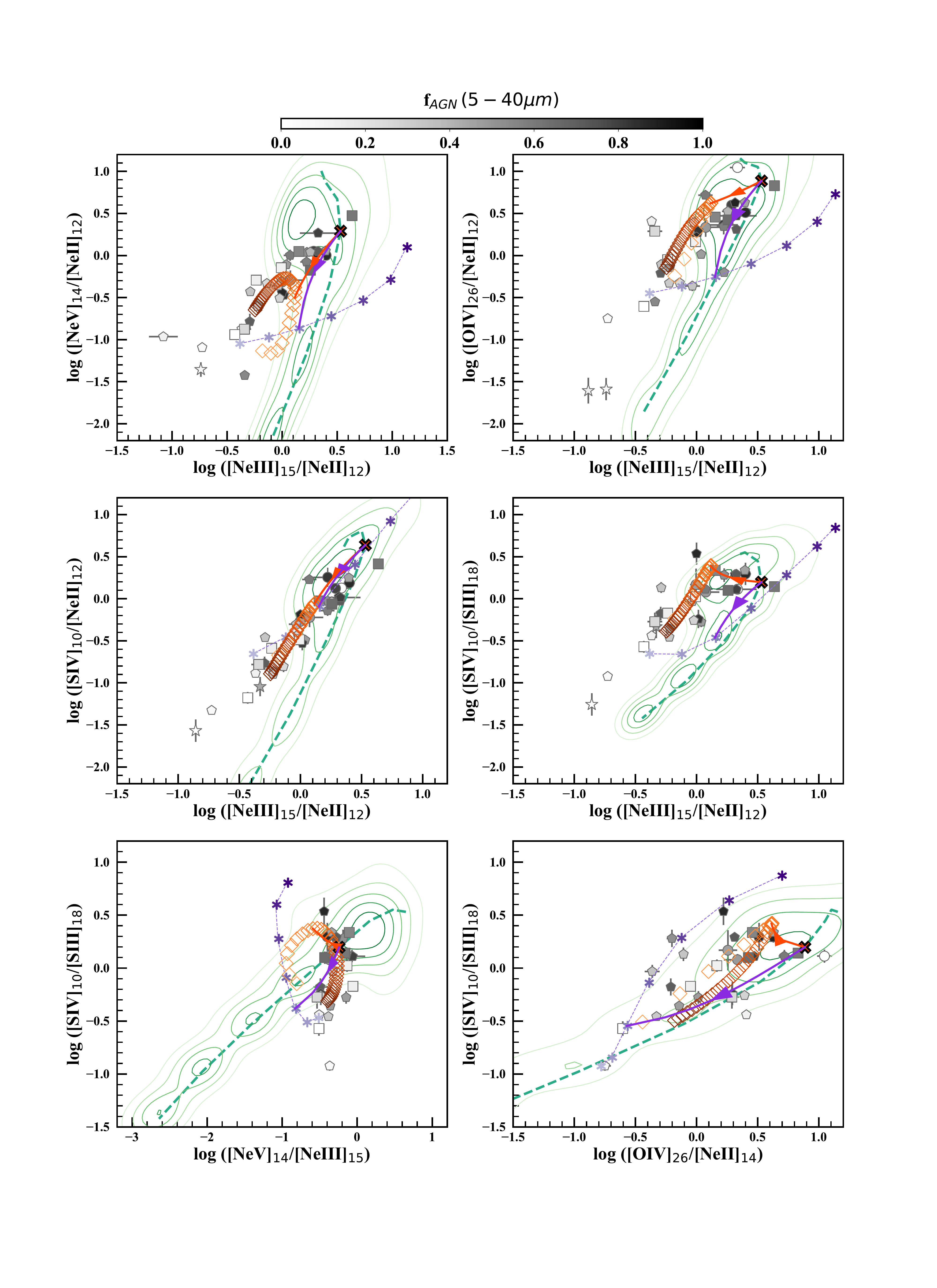}
     \caption{Mid-IR line ratio diagrams. Symbols and colors of the observations and models are the same as in Fig. \ref{fig:BPT}. The green contours show the full grid of the AGN models from F16 with $Z>1/3Z_{\odot}$ described in Sect. \ref{sec:photomod}.}
     \label{fig:mir_diagn}
\end{figure*}

We started with the \nev/\neiii\ versus \neiii/\neii\ (top-left panel of Fig. \ref{fig:mir_diagn}), which is only slightly sensitive to the ionization state of the gas \citep[as detailed in][]{Groves2006b} and insensitive to metal abundance because based only on Ne-lines. Irrespective of the Seyfert type, the \nev/\neii\ ratio is lower for those targets with a known strong contribution from star formation, as traced for example by lower fractional contribution of the AGN to the mid-IR continuum at 5-40 \mum, f$_{\rm AGN}$. The trend is similar when considering \oiv/\neii\ (top-right panel of Fig. \ref{fig:mir_diagn}; see also Fig. 11 of G16). This is because ionization from star-forming regions contributes primarily to the \neii\ line, with the effect of reducing the \nev/\neii\ and \oiv/\neii\ ratios. These ratios are useful diagnostics of the radiation field, because of the different ionization potentials required to produce Ne$^{4+}$ and O$^{3+}$  (97.1 and 54.9 eV, respectively) and Ne$^{+}$ (21.6 eV). We note that \cite{Inami2013} used a value of \nev/\neii\ $\geq 0.1$ to classify their sources as AGN-dominated in the mid-IR. We briefly illustrate how observations of star-forming galaxies from the literature compare with this threshold value in Appendix \ref{app:midIR_full}. Mrk\,897, the only non-Seyfert of our sample with all three neon lines detected, shows line ratios consistent with those of starburst galaxies. The two intermediate Seyferts, NGC\,4602 and NGC\,7469, which have values of \nev/\neiii\ around 0.1, have a fractional contribution of the AGN to the mid-IR of 12 and 5\%, respectively. Interestingly, the intermediate Seyfert NGC\,7603, which occupies the mixed-region of the BPT \citep*{BPT} diagram (Fig. \ref{fig:BPT}), has \nev/\neiii\ $<0.1$, close to those of starburst galaxies.
 
We then turn to line ratios based on emission lines of lower ionization potential than \nev, where the contribution from star formation can be more significant. 
The \siv/\neii, \neiii/\neii\ and \siv/\siii\ ratios increase with the AGN contribution f$_{\rm AGN}$ as shown in the two central panels of Fig. \ref{fig:mir_diagn}. 
The \siv/\neii\ versus \neiii/\neii\ diagram has been used as primary diagnostics by \cite{Inami2013} to analyze a sample of 202 local luminous IR galaxies. The sources in our sample have $1.5 < $ log(\siv/\neii) $<0.5$ and  $-1.0 <$ log(\neiii/\neii)  $< 0.5 $, similarly to the AGN-dominated sources of the \cite{Inami2013} sample (defined to have \nev/\neii\ $\geq 0.1$). The two targets with the lowest \siv/\neii\ and \neiii/\neii\ ratios, NGC\,7469 and NGC\,7496, have f$_{\rm AGN} \approx 0$ and lie in the area occupied by the star-formation-dominated sources of \cite{Inami2013} and the starburst galaxies collected by \citet[][see their Sect. 2.2]{Fernandez-Ontiveros2016}, as shown in Fig. \ref{fig:mir_diagn_appendix}. 
Similarly, in the \siv/\siii\ versus \neiii/\neii\ diagram, the ratios decrease with decreasing f$_{\rm AGN}$. All but three of the 19 targets with f$_{\rm AGN} < 0.4$ are starburst galaxies based on the distance from the star-forming main sequence of \cite{Bluck2020}. These objects with low f$_{\rm AGN}$ have values of these line ratios close to those of starburst galaxies, indicating that star formation contributes significantly to the line emission  (central-right panel of Fig. \ref{fig:mir_diagn_appendix}). 

We now explore the \siv/\siii\ versus \nev/\neiii\ diagram (bottom-left panel of Fig.  \ref{fig:mir_diagn}). While \siv/\siii\  is sensitive to the AGN dominance, f$_{\rm AGN}$, \nev/\neiii\ is mainly sensitive to the hardness of the radiation field, irrespective of the relative contributions of different ionizing sources to the mid-IR emission. This is because the \nev\ and \neiii\ lines have high-ionization energies ($\gtrsim 40$ eV) and are primarily dominated by the AGN (though a contribution from star formation can be present in the \neiii\ emission). The \oiv/\neiii\ ratio (not shown), similarly to \nev/\neiii, does not show any strong trend with f$_{\rm AGN}$, while the \oiv/\neii\ ratio (bottom-right panel of Fig. \ref{fig:mir_diagn}) increases with increasing f$_{\rm AGN}$, as in the case of the \nev/\neii\ ratio (top-left panel). The \nev/\neiii\ and  \oiv/\neii\ ratios are overall higher than those observed in starburst galaxies. This is because our targets have been originally selected as AGN candidates based on strong optical line emission.

\cite{Richardson2022} explored other mid-IR diagnostic diagrams that are different from those in the current section, namely \neiii/\neii\ versus \oiv/\neiii, \oiv/\neiii\ versus \siv/\neii\ and \oiv/\siii\ versus \siv/[\ion{Ar}{ii}]6.9\mum\ and proposed demarcation lines on these same diagrams for separating AGN and star formation. We checked the mid-IR data of our sample, and our targets classified as Seyfert have $log_{10}$\oiv/\neiii$>-0.85$ and $log_{10}$\oiv/\siii$> -0.34,$ falling therefore in the AGN-dominated regions of the diagnostics discussed in Sect. 4.2 of \cite{Richardson2022}. Two of the non-Sy, Mrk\,897 and NGC\,6810, have $log_{10}$\oiv/\siii of $-1.38$ and $-1.2$, respectively and they fall in the AGN-dominated area of the first two diagrams (i.e., \neiii/\neii\ versus \oiv/\neiii, \oiv/\neiii\ versus \siv/\neii) while in the composite (AGN and star formation) region in the \oiv/\siii\ versus \siv/[\ion{Ar}{ii}]6.9\mum\ diagram.

\section{Theoretical optical and IR line emission} \label{sec:model_comparison}

We now compare our optical and mid-IR data measurements with theoretical predictions from emission line models described in Sect. \ref{sec:photomod}. Specifically, we first explore AGN photoionization models (Sects. \ref{sec:AGNcomparison} and  \ref{sec:agnmodels}) and then investigate how additional contributions from star formation (Sect. \ref{sec:sf}) and shocks (Sect. \ref{sec:shock}) can impact the line ratios. We also investigate how combined optical and mid-IR line ratio diagrams can help in the identification of the different ionizing sources in Sect. \ref{sec:mixed}.

\subsection{Models of nebular emission from different ionizing sources}\label{sec:photomod}

We explored models of the emission from the gas in the NLRs of AGN computed by \citet[hereafter F16]{Feltre2016}, with some updates, as explained in Sect. 4.1 of \cite{Mignoli2019}. The calculations have been updated using the photoionization code \textsc{CLOUDY} \citep[c17.02;][]{Ferland2017}. The shape of the ionizing radiation field is represented by a broken power law, with the UV spectral index $\alpha$ ($F_{\nu}\propto \nu^{\alpha}$ between 5 and 1000 eV) as variable parameter, ranging from -1.2 to -2.0 \citep[see also][]{Groves2004}. Additionally, these models are parametrized in terms of other physical quantities (see Table 1 of F16), such as the volume-averaged ionization parameter, \ionpar\ -- that is, the dimensionless ratio between the number density of ionizing photons and that of atoms of neutral hydrogen -- the hydrogen density of the gas cloud, $n_{\rm H}$, the gas metallicity, $Z$, and the dust-to-metal mass ratio, $\xi_{\rm d}$ (which accounts for the depletion of refractory metals onto dust grains).

In addition to the AGN-driven ionization, we consider the potential contributions from star formation and fast radiative shocks to the line emitted spectra. 
For the star formation component, we used models of the nebular emission from gas ionized by single, young and massive stars developed by \cite{Gutkin2016}, using the version c13.03 of \textsc{CLOUDY} \citep{Ferland2013}. These models, hereafter star-forming galaxy (SF) models, describe the nebular emission from the gas in spherical \ion{H}{ii} regions using the latest version of the stellar population synthesis models of \citet[][Charlot \& Bruzual in preparation]{Bruzual2003}. These models incorporate updated stellar evolutionary tracks \citep{Bressan2012}, including new prescriptions for the evolution of the most massive stars ($\gtrsim 25 M_{\odot}$, Wolf-Rayet phase, \citealp{Chen2015}; see also Appendix A of \citealp{Plat2019}). The adjustable parameters are the volume-average ionization parameter, \ionpar$^{\star}$, the hydrogen density of the gas cloud, $n_{\rm H}^{\star}$, the interstellar gas metallicity, $Z^{\star}$, and dust-to-metal mass ratio, $\xi_{\rm d}^{\star}$ \citep[see Table 1 of][]{Gutkin2016}. The settings for metal abundances and depletion factors used to compute the AGN and SF models are the same. The SF models from \cite{Gutkin2016} provide the nebular emission from a whole galaxy, parametrized in terms of ``galaxy-wide'' parameters, by convolving the spectral evolution of single, ionization bounded  \ion{H}{ii} regions with a constant star formation history. As reference comparison with AGN models, throughout this work, we assume star formation at a constant rate for 10 Myr. Since most of the ionizing photons are released at ages less than 10 Myr by a single stellar generation \citep{Charlot1993, Binette1994}, this is a sufficient time to reach a steady population of \ion{H}{ii} regions. 

An additional adjustable parameter of the models is the carbon-to-oxygen abundance ratio ($C/O$). In this work we keep C/O fixed to the solar value \citep[$C/O_{\odot} = 0.44$; Sect. 2.3.1 of][]{Gutkin2016} as we do not investigate any carbon feature that could help constrain this parameter \citep[for a discussion on the C/O abundance ratio in AGN, see][]{Nakajima2018}. The \cite{Gutkin2016} and F16 models include dust physics \citep[e.g.,][for grain physics in \textsc{CLOUDY}]{vanHoof2004} and a self-consistent treatment of metal abundances and depletion onto dust grains. 
We note that we express the ionization parameter as the volume-averaged value defined in \citet[][see also footnote 7 of \citealt{Hirschmann2017}]{Plat2019} following the definition in Eq. (B.6) of \cite{Panuzzo2003}, which is a factor of 9/4 larger than the ionization parameter at the Strömgren radius used by F16 and \cite{Gutkin2016}.
We refer to Table 1 in F16, Table 3 in \cite{Gutkin2016} and to the following Sects. \ref{sec:AGNcomparison} and \ref{sec:sf} for the range of the parameter values of the AGN and SF models.

The UV radiation originated by fast radiative shocks can contribute to the line emitted spectra. We investigate the contribution from shock-ionized gas considering the recent models by \cite{Alarie2019}. This model grid has been computed using the \texttt{MAPPINGS V} shock and photoionization code \citep{Sutherland2017} and is publicly available from the 3MdBs database\footnote{\url{http://3mdb.astro.unam.mx:3686/}}. This database includes models with the same sets of element abundances as those adopted in the stellar and AGN photoionization models described above, although metal depletion onto dust grains is not included in the case of fast radiative shocks as grain-grain collisions and thermal sputtering can efficiently destroy dust \citep[][and their Sect. 2 for a discussion about the effect of dust depletion on the output spectra]{Allen2008}. 
Other main adjustable parameters \citep[see, e.g.,][for more details]{Allen2008, Alarie2019} are the shock velocity, $v^{\rm sh}$ (from 10$^{2}$ to 10$^{3}$ km s$^{-1}$), the pre-shock density, $n_{\rm H}^{\rm sh}$ (from 1 to 10$^{4}$ cm$^{-3}$) and the transverse magnetic field, $B$ (from 10$^{-4}$ to $10$ $\mu$G). 
In the following sections, we investigate predictions of photoionization models of AGN, star-forming regions and shocks in diagnostic diagrams based on mid-IR lines. We refer to the work of \cite{Alarie2019} for the analysis of shock models in optical diagnostic diagrams and to \cite{Plat2019} for a study of the combined effect of ionization from \ion{H}{ii} regions, AGN and shocks on optical and UV line ratios.

\subsection{Comparison with nebular emission from AGN}\label{sec:AGNcomparison}

The suite of AGN models described in Sect. \ref{sec:photomod} reproduces well the observed optical line ratios of our Seyfert galaxies as illustrated in Fig. \ref{fig:BPT}.
For illustration purposes, in Fig. \ref{fig:BPT}, we show photoionization models of the NLRs in AGN (dashed lines) with hydrogen density $n_{\rm H}$ of $10^3$ cm$^{-3}$,  UV spectral index $\alpha$ of $-1.7$,  $\xi_{\rm d}$ of 0.3, and volume-averaged ionization parameter increasing from \logU = -4.5 to -1.5 from bottom to top. Models with gas metallicity close to the solar value, $Z=0.017$, are shown in each panel, while the left panel displays models with metallicities $0.5 Z_{\odot}$ and $1.5 Z_{\odot}$ (lighter and darker dashed lines). For the exploration of the whole parameter space of these photoionization models in optical diagnostic diagrams, we refer to the original work of F16.
  
We now investigate how these models populate mid-IR line-ratio diagrams. The light green contours in Fig. \ref{fig:mir_diagn} represent the whole suite of models from F16. The dashed green line represents models for different values of ionization parameter (\logU\ from $-4.5$ to $-1.5$, increasing from bottom to top) and other parameters fixed, $\alpha=-1.7$, $\xi_{\rm d}=0.3$, n$_{\rm H}=10^3$ cm$^{-3}$ and $Z=0.017$ Z$_{\rm \odot}$ (as shown in Fig.  \ref{fig:BPT}). The green-shaded contours illustrate the predictions from our suite of AGN models with metallicity higher than $1/3 \, Z_{\odot}$ (see Table 1 of F16 for the full range of values). The choice to limit the comparison to models with $Z > 1/3 \, Z_{\odot}$ is supported by various observational evidence, as detailed further in Sect. \ref{sec:metal} (e.g., top left panel of Fig. \ref{fig:mir_diagn_AGNmod}). 

While the mid-IR line ratios of our targets with $f_{\rm AGN} \gtrsim 0.5$ are well reproduced by the AGN photoionization models, these same models fail to account simultaneously for more than one of the mid-IR line ratios measured in the spectra of the targets with lower values of $f_{\rm AGN}$.
For example, if we take the \nev/\neii\ and \neiii/\neii\  ratios separately, the AGN model grid predicts values that are in the range of those measured in the mid-IR spectra of our sample (top-right panel of Fig. \ref{fig:mir_diagn}). When combined together, these ratios are not reproduced simultaneously by the AGN model grid, even considering the entire range of parameters explored in F16. 

This suggests that further modeling, including improvements on the AGN models and/or the inclusion of additional contributions from other ionization sources, is needed to fully reproduce the observations. We investigate this by analyzing the impact of extending the parameter space of the AGN grid, as detailed in Sect. \ref{sec:agnmodels}, and then considering additional sources of ionizing photons from either star formation or shocks, as described in Sects. \ref{sec:sf} and \ref{sec:shock}. 

\subsection{Exploring AGN photoionization models}\label{sec:agnmodels}

\subsubsection{Metallicity}\label{sec:metal}
We first note that models with $Z<1/3\,Z_{\odot}$ can help explain some (but not all) the combinations of mid-IR line ratios, as shown, for example, in the top-central versus the top-right panel of Fig. \ref{fig:mir_diagn_AGNmod}. However, we do not favor models with such a low metallicity for our sample of local Seyfert. This is because, first, the models with $Z<1/3\,Z_{\odot}$ do not cover the area occupied by our observations in optical diagnostic diagrams (top-left panel of Fig. \ref{fig:mir_diagn_AGNmod}) but move toward the left side of the BPT diagrams populated by \hii\ regions  \citep[e.g.,][]{Groves2004,Feltre2016,Hirschmann2017}. 
In addition, we used the  \cite{Storchi-Bergmann1998} calibration (see their Eq. 2) to compute the metallicity of those targets with both the \oiii/\hb\ and \nii/\ha\ measurements available, and obtained oxygen abundances in the range $8.33 < 12+log(O/H) < 9.22$. The minimal value of 8.22 for $12+log(O/H)$ corresponds to $\approx 1/3$ the solar value of the gas-phase oxygen abundance adopted in the AGN models,namely, $12 + log(O/H)=8.68$ for $Z_{\odot} = 0.01524$ and $\xi_{d,\odot}  = 0.36$. 

The range of stellar masses of the host galaxies of our targets is $ 9.0 <log(M_{\star} / M_{\odot})< 11.5$ (mean value of 10.5), as derived from the SED fitting described in G16. Adopting the mass-metallicity relation of local galaxies \citep{Thomas2019}, our targets with stellar masses $\gtrsim10^{10} M_{\odot}$ (lower limit of validity of the relation) are unlikely to have metallicity below half the solar value. We obtain the same result, if we consider the analysis by \cite{Dors2020}, where no trend between stellar mass and metallicity was found for local galaxies in the mass range $9.4 <log(M_{\star} / M_{\odot})< 11.6$, irrespective of the method used to infer the metal content. For the lowest stellar masses of our sample ($\lesssim10^{9.5} M_{\odot}$, 4 of our targets), the mass-metallicity relation of star-forming galaxies \cite[e.g.,][and references therein]{Lequeux1979, Tremonti2004,Curti2020} suggests oxygen abundances higher than $\approx 1/2$ the solar value adopted in our models. We acknowledge that for the lower stellar masses ($\lesssim10^{9.5-10} M_{\odot}$) there is a larger uncertainty in the observed metallicity at a given mass, meaning that this consideration alone is not sufficient for ruling out photoionization models with $Z<1/3\, Z_{\odot}$.

\subsubsection{Density}
\begin{figure*}
\centering
  \includegraphics[width=17cm]{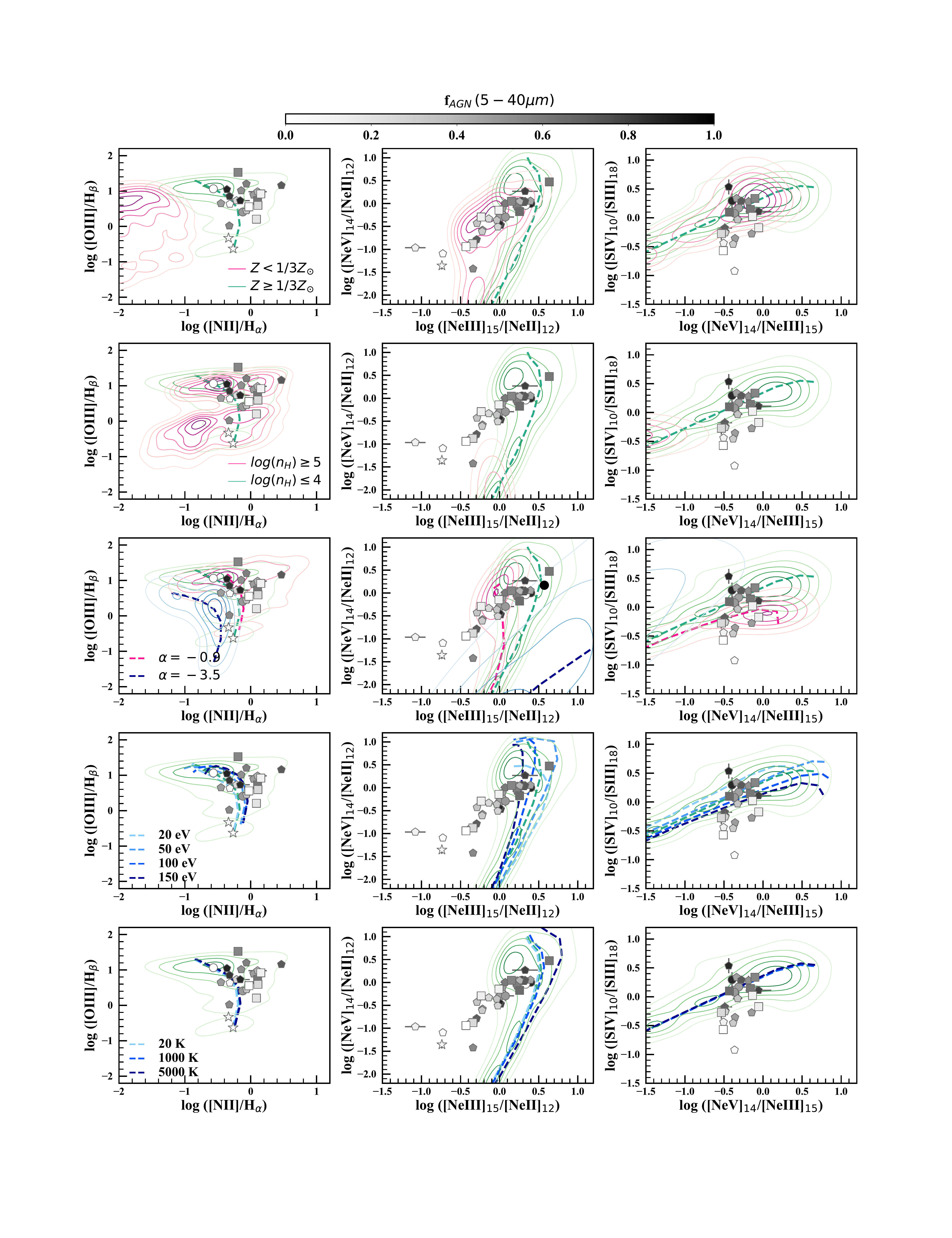}
     \caption{Examples of one optical (\oiii/\hb\ versus \nii/\ha) and two mid-IR (\nev/\neii\ versus \neiii/\neii\ and \siv/\siii\ versus \nev/\neiii) line ratio diagrams. Symbols and colors of the observations are the same as in Fig. \ref{fig:BPT}. The dashed green lines and contours have the same meaning as in Fig. \ref{fig:mir_diagn}. Magenta contours in the first and second row of panels show AGN models with $Z<1/3 Z_{\odot}$ and $n_{\rm H}\geq10^{5}$ cm$^{-3}$, respectively. Dashed blue and magenta curves and contours in the third row are models with $\alpha=-0.9$ and $\alpha=-3.5$ and with the other parameters the same as the green dashed curves and contours. The dashed blue shaded curves in the last two rows show AGN models with different energy peaks of the Big Blue Bump (second last row) and with the lower-limit temperature set to stop the calculations (last row), with values as labeled in the left panels.}
     \label{fig:mir_diagn_AGNmod}
\end{figure*}

Given that ratios of two high-ionization emission lines, like \nevrl/\nevl, can trace densities as high as $10^{5-6}$ cm$^{-3}$ and for consistency with previous works \citep[e.g.,][]{Spinoglio2015, Fernandez-Ontiveros2016}, we extended our suite of AGN photoionization calculations toward higher hydrogen densities, with n$_{\rm H}=10^5$ and $10^6$ cm$^{-3}$.
  While models with n$_{\rm H}\geq10^5$ cm$^{-3}$ predict line ratios still in marginal agreement with observations in the optical, these are by orders of magnitude different from what is observed in the mid-IR (second row of panels from top in Fig. \ref{fig:mir_diagn_AGNmod}). The inappropriateness of models with extreme values of hydrogen density to reproduce the observed line ratios is consistent with the conclusions reached by other authors in the analysis of NLR emission in optical and UV wavelengths \citep[e.g.,][]{Nagao2006,Feltre2016}. Specifically, \nev/\neii\ predicted by the models with n$_{\rm H}>10^4$ cm$^{-3}$ is too low compared to \neiii/\neii\ and \siv/\siii\ to explain the objects with low f$_{\rm AGN}$. The situation is similar for \nev/\neiii\ and \oiv/\neii\ (not shown). This is because, at fixed other parameters, increasing the hydrogen density raises the dust optical depth. This implies extra absorption of energetic photons and causes \nev/\neii, \nev/\neiii\ and \oiv/\neii\ to drop.
  Previous works from \cite{Spinoglio2015,Fernandez-Ontiveros2016}, including also part of our targets, have shown a stratification of densities with ratios, like \siiilred/\siiil, tracing lower density gas ($\approx 10-10^{3}$ cm$^{-3}$) than those based on higher-ionization lines, like \nevrl/\nevl, which, instead, arise from the innermost regions of the AGN NLRs ($\approx 10^{2}-10^{4}$ cm$^{-3}$). When both lines from the same transitions were available, we derived the electron density $n_{\rm e}$ for our targets using \textsc{PyNeb}, a \textsc{PYTHON} package for the analysis of emission lines \citep{Luridiana2015}. We employed the function {\itshape getTemDen}, assuming an electronic temperature of 10$^4$\,K, and find that the highest values of log($n_{\rm e}$) are 3.15 and 4.2, for \siiilred/\siiil\ and \nevrl/\nevl, respectively. These values are in agreement with those found in the comparison between photoionization models and observations (Fig. \ref{fig:mir_diagn_AGNmod}).

\subsubsection{Ionizing radiation field}\label{sec:ionrad}
We then considered models with different UV spectral indices, $\alpha$, namely a flatter (harder, $\alpha=-0.9$) and steeper (softer, $\alpha=-3.5$) ionizing continuum. The latter was used to represent the LINER emission \citep{Fernandez-Ontiveros2016}, while $\alpha=-0.9$ is close to the values of $-0.8$ and $-1.0$ used by other authors to interpret line ratios of Type 2 AGN \citep[e.g.,][]{Perez-Montero2019, Dors2020}.  
We refer to the Sect. \ref{sec:disc_ion} for a more detailed discussion on the parametrization and shape of the AGN ionizing radiation field and its impact in our study.

Since, $\alpha=-3.5$ is used to represent the low-ionization nuclear emission regions, one would expect a higher \neii\ (and, in turn, a lower \neiii/\neii) compared to models with lower $\alpha$, as those of of the grid described in Sect. \ref{sec:photomod}. This is what we observe in our computations (blue contours and dashed line in the third row of panels of Fig. \ref{fig:mir_diagn_AGNmod}). At the same time, the \nev\ line intensity of models with $\alpha=-3.5$ is not high enough to reproduce the observed data. In addition, even if models with $\alpha=-3.5$ could marginally explain some objects with low f$_{\rm AGN}$ in the \nev/\neiii $-$ \neiii/\neii\ plane, these models do not reproduce the corresponding optical data.  

Flattening the ionizing radiation from $\alpha=-2.0$ to $-0.9$ reduces the \nev/\neii, \neiii/\neii\ and \siv/\siii\ ratios and brings the models in better agreement with part of the Seyfert data with low f$_{\rm AGN}$ (pink contours and dashed line in the third row of panels of Fig. \ref{fig:mir_diagn_AGNmod}). This is because in models with flatter spectra, the emission from the lines with higher ionization energies, like \nev\ in the case of \nev/\neii\ or \siv\ in the case of the \siv/\siii\ ratio, is more compact around the central source (see the discussion in Sect. \ref{sec:disc_ion}). 

To investigate further the impact of the incident radiation field, we computed AGN NLR models using a different shape of the ionizing radiation. 
Specifically, we used the \texttt{AGN} command in \textsc{CLOUDY} \cite[Sect. 6.2 of ``Hazy 1'' documentation of c17.02][]{Ferland2017} varying the temperature of the peak of the AGN Big Blue Bump from 20 to 150 eV. A similar range of temperatures and their impact on optical line ratios have been explored by \cite{Thomas2016} using the \textsc{MAPPINGS} photoionization code \citep{Sutherland2018}. The other parameters of the \texttt{AGN} command are set to the default values of \textsc{CLOUDY}, namely $-1.4$ for the X-ray to UV ratio, $\alpha_{\rm OX}$ \citep{Zamorani1981}, $-0.5$ for the low-energy slope of the Big Blue Bump continuum, $\alpha_{\rm UV}$ \citep{Elvis1994, Francis1993}, $-1.0$ for the slope of the X-ray component. Shifting the energy peak from 20 to 150 eV, reduces the \neiii/\neii\ ratio but not enough to match the full set of observations of our sample (blue-shaded lines in the fourth row of panels of Fig. \ref{fig:mir_diagn_AGNmod}).

\subsubsection{Stopping criteria of the calculations}

The calculations by F16 are stopped either when the electron density falls below 1 per cent of $n_{\rm H}$ or when the temperature falls below 100 K (i.e., at the edge of the Str\"omgren sphere). 
To explore how the criteria for setting the spatial extent of the calculations impact the predictions of the line ratios, we removed the condition on the electron temperature and stopped the calculations at different temperatures, from 20 K to 5000 K. Setting 20 K as the stopping criterion enabled us to model the gas beyond the fully ionized region \citep[see][for models of SF galaxies]{Vidal-Garcia2017}, while allowing a minimum temperature of 1000K or, even more extreme, 5000K ensured we modeled only the emission of the photoionized gas \citep[with hydrogen almost completely ionized; see also Sect. 3.1.2 of][]{Dors2022}. 
As illustrated in Fig. \ref{fig:mir_diagn_AGNmod} (blue shaded lines in the second row of panels from the bottom), the model predictions are not significantly sensitive to the lower-limit temperature. This is because we are exploring high- and intermediate-ionization emission lines that arise in fully ionized regions \citep[see also Sect. 3.1 of][]{Nagao2006}. 
We obtain similar results when adopting different optical depths as criterion to stop the photoionization calculations (not shown).

\subsection{Contribution from star formation to line emission}\label{sec:sf}

An additional contribution from star formation to the AGN nebular emission can impact the total emitted spectra of the ionized gas \citep[e.g.,][]{Davies2016}. To investigate this, we consider models of nebular emission from star-forming galaxies by \cite{Gutkin2016} with ionization parameter in the range $-3.6 < log ( \langle U^{\star} \rangle ) < -0.65$, $Z^{\star}=0.0017$,  $n_{\rm H}^{\star} = 10^2$ cm$^{-3}$, and $\xi_{\rm d}^{\star}=0.3$. As illustrated in Fig. \ref{fig:mir_diagn}, we started from a ``reference'' AGN model with Z$=0.017$,  $log ( \langle U \rangle ) = -2.5$, $n_{\rm H}=10^3$cm$^{-3}$, $\xi_{\rm d}=0.3$ and $\alpha=-1.7$ (indicated with the black cross) and add a fractional contribution from star formation to the total (AGN+SF) \hb\ line emission (as done for Fig. \ref{fig:BPT}). The stars indicate predictions of combined AGN and SF models with 90\% contribution to the total \hb\ line from star formation. The different shades of the asterisk symbols indicate the variation of the ionization parameter of SF models (decreasing from the darkest to the lighter shade). The magenta line with the arrow indicates the effect of adding a 0 to 90\% contribution from star formation to the reference. AGN model considering a SF model with $Z^{\star}=0.017$, $log (\langle U^{\star} \rangle ) = -3.0$, $n_{\rm H}^{\star}=10^2$cm$^{-3}$ and $\xi_{\rm d}^{\star}=0.3$. Fig. \ref{fig:mir_diagn_appendix} in Appendix \ref{app:midIR_full} shows how the suite of SF models described in Sect. \ref{sec:photomod} populates the mid-IR line-ratio diagrams explored in this work. 

Adding a fractional contribution from 0 to 90\% of star formation to the AGN nebular emission explains the scatter observed in the \siv/\siii\ versus \oiv/\neii\ and  \siv/\siii\ versus \nev/\neiii\ diagrams (bottom panels of Fig. \ref{fig:mir_diagn}). We note that the contribution from star formation to the nebular emission lowers \nev/\neiii\ and \nev/\neii\ (or \oiv/\neiii\ and \oiv/\neii). This is because stars alone do not produce a significant amount of photons that are hard enough to ionize \nev\ and \oiv\ (see Fig. \ref{fig:ionspec}). Given that the continuous mid-IR emission of objects with f$_{\rm AGN} < 40\%$ is dominated by reprocessed stellar emission, one would expect a significant contribution from star formation to the line emission as well. 
We find that a fractional contribution from star formation can explain the observations of line ratios, such as \nev/\neii\ versus \neiii/\neii\ and \oiv/\neii, when considering stellar models with values of the ionization parameter $log \langle U  \rangle < -3$, which are common values for local star-forming galaxies \cite[e.g.,][]{Brinchmann2004}.

\subsection{Impact of fast shocks on line emission}\label{sec:shock}
 
 We investigate how fast radiative shock impact the line emitted spectra by combining the AGN models of F16 with a sub-grid of ``pure'' shock models by \cite{Alarie2019}. 
We note the choice to combine different models, rather than using a code that generates composite of shocks$+$AGN \citep[such as the SUMA code;][]{Viegas-Aldrovandi1989, Contini2019, Dors2021}, is for consistency with the previous sections and because the \cite{Alarie2019} models are already computed for the same sets of element abundances as those adopted in the stellar and AGN photoionization models.

We consider the full available range of shock velocities from the \cite{Alarie2019} grid, $v^{\rm sh}$ from 200 to 1000 km/s and, for simplicity, keeping fixed the other parameters to a given value, namely gas metallicity $Z^{\rm sh}=0.017$, pre-shock density $n_{\rm H}^{\rm sh} = 10^{2}$ cm$^{-3}$ and the transverse magnetic field, $B = 1 \mu$G. These combined models are shown in Fig. \ref{fig:BPT} (optical) and Fig. \ref{fig:mir_diagn} (mid-IR). The empty diamonds, color-coded based on the shock velocity (increasing from 200 to 1000 km/s from the lighter to the darker shade), indicate model predictions for a fractional contribution from shocks of 90\% to the total \hb\ flux. As in Sect. \ref{sec:sf}, this contribution has been added to a reference AGN model with Z$=0.017$,  $log ( \langle U \rangle ) = -2.5$, $n_{\rm H}=10^3$cm$^{-3}$, $\xi_{\rm d}=0.3$ and $\alpha=-1.7$.  
The solid line with the arrow shows the effect of adding a 0 to 90\% fractional contribution from shocks to the \hb\ flux of the aforementioned AGN model considering a shock model with velocity of $400$ km/s. 
Exploring values of relative contribution from shocks up to 90\% is supported by identification of possibly shock-dominated galaxies in the analysis of JWST NIRSpec spectrograph observations \citep{Jakobsen2022, Ferruit2022} by \cite{Brinchmann2022}. Moreover, \cite{Contini2004} found that the IR emission of some AGN can be explained primarily through shock models.

Starting with any of the AGN models (green contours), the addition of a shock component can bring the models in agreement with the observed data points. 
We note that the addition of a contribution from shocks to the AGN nebular emission allows us to account for the observed trends in all the mid-IR ratios investigated in this work. 
This can be fully appreciated in the figures in Appendix \ref{app:midIR_full} showing the regions of the diagrams populated by pure (i.e. 100\% contribution) shock$+$precursor models (orange contours). 

\subsection{Combining optical and mid-IR line ratios}\label{sec:mixed}

\begin{figure*}
\centering
  \includegraphics[width=17cm]{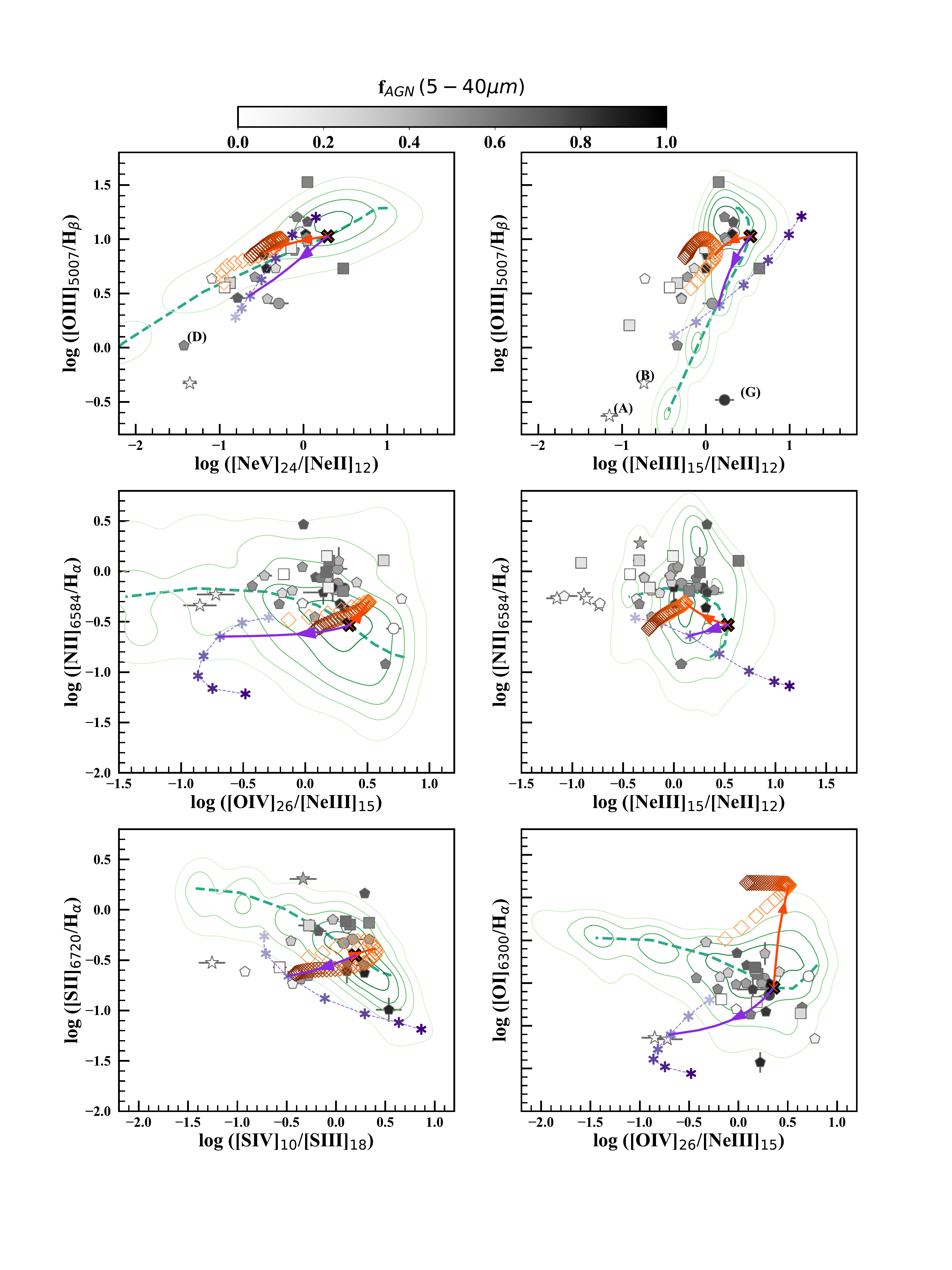}
     \caption{Examples of diagrams involving one optical and one mid-IR ratio. Symbols and colors of the observations and models are the same as in Fig. \ref{fig:mir_diagn}.}
     \label{fig:mir_opt_diagn}
\end{figure*}

We explore how AGN photoionization models can best reproduce combinations of optical (\oiii/\hb, \nii/\ha, \sii/\ha,\ and \oi/\hb) and mid-IR  (\siv/\siii, \nev/\neiii, \nev/\neii, \neiii/\neii, \oiv/\neii,\ and \oiv/\neii) line ratios. While in this section we analyze all the possible combinations of an optical and a mid-IR ratio among those listed above, for the sake of simplicity, we show only some illustrative diagrams in Fig. \ref{fig:mir_opt_diagn}. The full set of diagrams investigated in the analysis can be found in Appendix \ref{app:midIR_full}.

We start considering diagnostics based on \oiii/\hb\ and the  \nev/\neiii, \nev/\neii, \oiv/\neii, \oiv/\neiii,\ and \siv/\siii\ ratios. Overall, we find that the combinations of these ratios are well reproduced by the AGN model grid described in Sect. \ref{sec:photomod}. A few sources have an \oiii/\hb\ too low (for a given value of the mid-IR ratio) to be explained by AGN models alone, but can be explained by adding a star formation component to the total emitted spectra (e.g., top-left panels of Fig. \ref{fig:mir_opt_diagn}). Two are the non-Sy CGCG381-051 and Mrk\,0897 (flagged with (A) and (B), respectively, in the figure) whose line ratios are compatible with star formation. Another is the intermediate Sy NGC\,7603 (flagged with (D) in the figure), which lies in the composite area of the BPT, as discussed in Sect. \ref{sec:bpt}. As illustrated by the purple arrow in Fig. \ref{fig:mir_opt_diagn}, the addition of a contribution from a nebular component of stellar origin reduces both the \oiii/\hb\ and the mid-IR ratios of interest, like \nev/\neii\ for example (top-left panel). Another outlier from the AGN model grid is the Sy 1 IRASF\,03450+0055 (flagged with (G) in the figure), whose optical line fluxes have been difficult to measure as the \oiii\ line is blended with the \fevii\ feature and \ha\ with \nii. 
This target is likely to be placed in the area of the diagrams occupied by the AGN when more secure measurements of the optical lines will be available. 

Our AGN model grid fails to reproduce the objects with the lowest f$_{\rm AGN}$ in the \oiii/\hb-\neiii/\neii\ plane, unless one considers models with very high density ($n_{\rm H} > 10^{5}$ cm$^{-3}$, not shown). However, we have already noted in Sect. \ref{sec:agnmodels} that models with such high densities do not reproduce other ratios, like \nev/\neii\ for example. As illustrated by the arrows in the top-right panel of Fig. \ref{fig:mir_opt_diagn}, considering an additional nebular component to the AGN emission, either from shocks or star formation, brings the models in agreement with observations. 

When considering the \nii/\ha\ or \sii/\ha\ versus the mid-IR ratios, we obtain similar results as those just described above. A few examples are shown in the central and bottom-right panels of Fig.  \ref{fig:mir_opt_diagn}. The AGN model predictions compare well with observations when the optical \nii\ha\ and \sii/\ha\ ratios are combined with \siv/\siii,  \nev/\neiii\  and \oiv/\neii. When combined with \neiii/\neii, the AGN emission needs an additional component to explain the observed ratios of the targets with low f$_{\rm AGN}$. 
The intermediate Sy MCG-03-34-064 has high values of \sii/\ha\ and \nii/\ha, while the non-Sy NGC\,7213 has high \sii/\ha. This is likely due to the difficulties in measuring partially blended features (\nii$+$\ha\ and the \sii\ doublet) in their optical spectra.  Our AGN grid only agrees well for targets in the \oi/\ha\ versus the mid-IR ratios planes, with the exceptions of two non-Sy (Mrk\,897 and NGC\,6810), whose emission is explained well by star formation (e.g., bottom-right panel of Fig. \ref{fig:mir_opt_appendix}).

While the observed data points, particularly for the targets with high f$_{\rm AGN}$, are centered on the bulk of the AGN models in most of the diagrams (e.g., \nii/\ha\ versus \neiii/\neii, \oi/\ha\ versus \oiv/\neiii), this is not true for the diagrams involving the \nev/\neiii\ and \nev/\neii\ ratios (e.g., central-left panel of Fig. \ref{fig:mir_opt_diagn}). This suggests that likely the same pure AGN model does not simultaneously reproduce the entire set of observed ratios. Observations are compatible with a combined models consisting of multiple nebular components (AGN, star formation and shocks).

\section{Discussion}\label{sec:discussion}

We discuss our findings in relation to the incident spectra adopted in the AGN models (Sect.\ref{sec:disc_ion}), the observed trends with the AGN fractional contribution to the mid-IR (Sect.\ref{sec:fagn}) and the need of additional ionizing sources to fully explain the observational data (Sect. \ref{sec:additional_sources}). We conclude the section discussing some caveats of our spectroscopic analysis (Sect. \ref{sec:caveats}).

\subsection{Incident spectra of AGN radiation}\label{sec:disc_ion}

The relative strengths of lines emitted in the NLRs of AGN strongly depends on the shape of the incident radiation. The AGN radiation from the accretion disk in the original grid of the F16 models is represented by a series of broken power laws, where the power law slope $\alpha$ (defined as $F_{\nu} \propto \nu^{\alpha}$) at the shorter wavelengths ($\lambda < 0.25 \, \mu$m) is a free parameter comprised between $-2.0$ and $-1.2$. 
This range encompasses the values of $\alpha$ at wavelengths shortward of the Ly$\alpha$ that have been empirically derived in samples of quasars (see below) and these are usually adopted as ``standard'' values in the literature \citep[e.g.,][]{Groves2004}.

Specifically, \cite{Zamorani1981} observed, in a sample of 79 X-ray detected quasars, a range of power-law slopes between optical and X-ray bands from $-0.91$ to $-1.87$, 
with a median value of $\approx$ -1.4. \cite{Zheng1997} found a power-law index in the extreme UV between 350 and 1050\,\AA\ of $-2.2$ and $-1.8$ for radio-loud and quiet quasars, respectively, and $-1.96$ for the full sample. They also noted a steepening of the continuum from 2200\,\AA\ to 1050\,\AA\ that could be modeled with a steeper spectrum, $\alpha=-0.99$ (close to the $-0.9$ used in Sect. \ref{sec:disc_ion}, with the difference that we adopted this slope down to shorter wavelengths). In a study of radio-intermediate and radio-loud quasars,  \cite{Miller2011} derived $\alpha$ in a  range between $\approx -0.8$ and $\approx-2.0$.  More recently, \cite{Zhu2019} measured values in the range $-1.4 < \alpha < -1.2$ in 15 radio quasars at $z>4$. 

Many authors developed models assuming the AGN incident spectrum from the \texttt{AGN} command in \textsc{CLOUDY} (as done in Sect. \ref{sec:agnmodels}), either keeping the X-ray to UV ratio fixed to a given value \citep[e.g., to the default value of $-1.4$ as in][]{Dors2019} or varying the slope among the values mentioned above \citep[e.g.,][]{Nagao2006, Dors2017,Humphrey2019}.  
\cite{Perez-Montero2019} investigated values as high as $-0.8,$ which were found to be the most adequate to reproduce the \oiii/\hb\ ratio of most Type 2 AGN by \cite{Dors2017}.  We note that the incident spectrum of the \texttt{AGN} command in \textsc{CLOUDY} contains an emission component, in addition to that of the Big Blue Bump, associated with the nonthermal X-ray radiation. \cite{Thomas2016,Thomas2018} built-up models using \texttt{MAPPINGS} and modeling an ionizing radiation field comprising two emission components: a pseudo-thermal accretion disk and a high-energy nonthermal component due to inverse Compton. 
The incident radiation field of the F16 models is meant to represent the pure thermal emission from the AGN accretion disk and does not include a nonthermal component. We acknowledge that various incident SEDs could affect differently the predicted line intensities, but a comprehensive comparison between the different parametrization choices of the AGN radiation are out of the scope of this work. A simple representation of AGN incident emission with a single-power law at the high energies, as in F16, enables straightforward studies of the impact of the steepness of the ionizing radiation on the predicted line emitted spectra. A more detailed comparison of the different shapes of AGN incident radiation field used in the models available in the literature will be presented in a theoretical study of optical emission lines by \cite{Vidal-Garcia2022}.

Turning the discussion on our results of Sect. \ref{sec:ionrad}, if we had to consider pure AGN photoionization, we find that we need an extended range of $\alpha$ values to explain the distribution of the observed line ratios in the diagrams of Figs. \ref{fig:mir_diagn} and \ref{fig:mir_opt_diagn}. 
In particular, all the optical line ratios slightly increase with the hardness of the radiation field with a flattening of the \oiii/\hb\ ratio for the highest $\alpha$. As noted above, a steeper ionizing spectrum helps reproduce the mid-IR line ratios of Fig. \ref{fig:mir_diagn_AGNmod}, with the exception of a few objects with low $f_{\rm AGN}$ whose mid-IR ratios are still not explained by the pure AGN models presented in this work. 
In these cases, the presence of shock-ionization could contribute in steepening the spectra, explaining the discrepancy between the $\alpha$ values from the observations and those inferred from the models \citep[e.g.,][]{Contini2019, Dors2022}. Another factor that could steepen the AGN ionizing spectra is the filtering of the AGN radiation through a gas screen in a wind-blown bubble or in the host galaxy \citep{Humphrey2019}. In this case, one should explore more complex composite models that account for the impact of the escape of AGN ionizing photons from the NLRs toward the ISM of the host galaxy on the total line emission spectra. 
Moreover, the influence of massive starburst relative to the AGN, as also suggested by a strong contribution of the star formation to the mid-IR continuum (i.e.,  low f$_{\rm AGN}$), could contribute to explain the observed line ratios. This is in line with our findings, namely that combinations of AGN and star formation nebular emission models can explain the full range of observed mid-IR line ratios. 

Mid-IR ratios like \nev/\neii\, \siv/\neiii,\ and \siv/\siii\ sightly decrease (up to 0.3 dex) with increasing $\alpha$ (i.e., with the hardening of the ionizing radiation field), while the \oiv/\neii, \neiii/\neii\ and \siv/\neii\ show a decrease of about 0.5 dex up to one order of magnitude. The \nev/\neiii\ and \oiv/\neii\ increase by about 0.3-0.5 dex with the hardening of the ionizing radiation.  
Given that \nev\ and \oiv\ have the highest ionization potential among the lines considered here and \neiii\ and \siv\ have a ionization potential higher than \neii\ (see Fig. \ref{fig:ionspec}), one would have expected the ratios between a line with a higher and another with a lower ionization energy to rise. This depends also on other model parameters such as metallicity and the ionization parameter, similarly to what is discussed in Sect. 4.2 of F16 for UV line ratios like \ion{C}{iv}$\lambda1550$/\ion{C}{iii}]$\lambda1908$.  
To summarize, we find that $\alpha$ has a significant impact on mid-IR line predictions, with variations up to one order of magnitude of mid-IR line ratios like \neiii/\neii\, rather than optical (with the exception of \oi). This suggests that the combination of observables in these two wavelength regimes along with analysis tools based on advanced statistical techniques, such as \textsc{BEAGLE} \citep{Chevallard&Charlot2016,Vidal-Garcia2022}, could provide valuable clues on the steepness of the ionizing radiation.

\subsection{Dependence on $f_{\rm AGN}$}\label{sec:fagn}

\begin{figure*}
\centering
  \includegraphics[width=19cm]{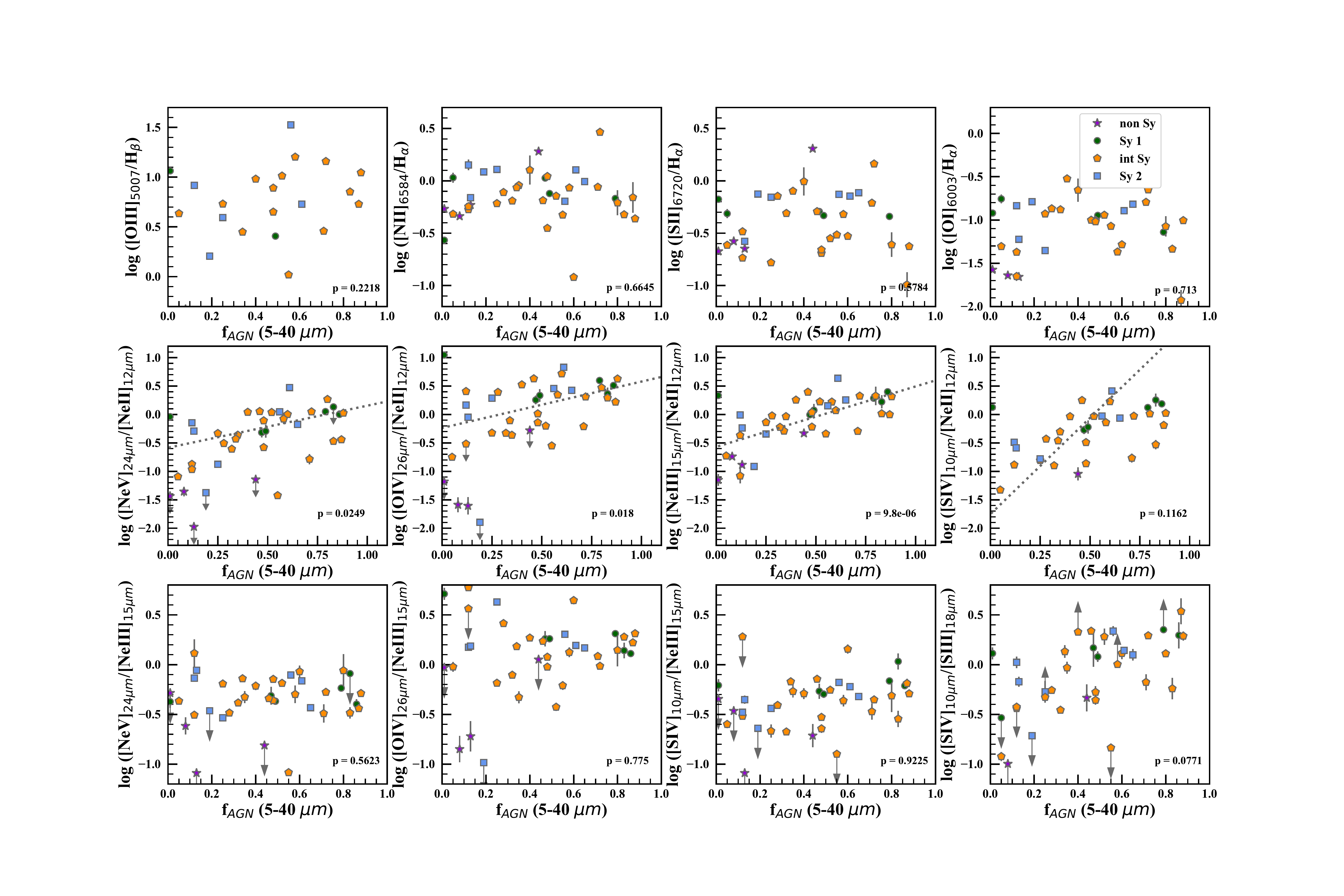}
     \caption{Optical (top panels) and mid-IR (central and bottom panels) line ratios as a function of the fractional contribution of the AGN to the mid-IR (5-40 \mum) continuum. Open symbols indicate the classification of the targets: green circles for Sy1, orange pentagons for int-Sy, light blue squares for Sy2, and violet stars for non-Sy. Upper and lower limits are marked with arrows. Each panel reports the p-value from the Spearman analysis at the bottom right. The dotted gray lines are the ODR fit results for the ratios with p-values smaller than 0.05.}
     \label{fig:ratios_fagn}
\end{figure*}

As mentioned above and shown in G16 (their figures 8 and 9), some of the mid-IR line ratios of interest here show a dependence on the fractional contribution of AGN to the 5-40 \mum\ mid-IR continuum, $f_{\rm AGN}$ (Fig. \ref{fig:mir_diagn}). We do not observe the same for optical line ratios, which populate the AGN-dominated areas of the BPT diagrams irrespective of $f_{\rm AGN}$ (Fig. \ref{fig:BPT}). This is better illustrated in Fig. \ref{fig:ratios_fagn}, which shows the observed optical and mid-IR line ratios, including upper and lower limits, versus $f_{\rm AGN}$ derived via SED fitting by G16. 

The optical line ratios all have a p-value from a Spearman rank analysis larger than $0.22$, meaning that any potential correlation is not statistically significant. The same is true for some mid-IR line ratios, particularly  \nev/\neiii, \oiv/\neiii,\ and \siv/\neiii, which all have a p-value larger than $0.5$. These mid-IR line ratios based on the ratio between a high- and an intermediate-ionization line are good tracer of the AGN activity, irrespective of the intensity of star formation. All the mid-IR line ratios between a high-ionization line and \neii\ show a statistically significant (p-value smaller than 0.05) positive correlation with $f_{\rm AGN}$. We performed a linear regression fit to these line ratios and $f_{\rm AGN}$ obtaining the following relations:

\begin{equation}
\begin{aligned}
log_{10}(\nev/\neii) = (0.74\pm0.39) \times f_{\rm AGN} + (-0.58\pm0.01)
\end{aligned}\label{eq:mir1}
,\end{equation}

\begin{equation}
\begin{aligned}
log_{10}(\oiv/\neii) = (0.81\pm0.39) \times f_{\rm AGN} + (-0.24 \pm 0.01)
\end{aligned}\label{eq:mir2}
,\end{equation}

\begin{equation}
\begin{aligned}
log_{10}(\neiii/\neii) = (1.06\pm0.65) \times f_{\rm AGN} + (-0.57 \pm 0.01)
\end{aligned}\label{eq:mir3}
.\end{equation}

When deriving the above relation, we considered the entire sample as we do not find specific sub-trends for one or more of the Seyfert types.
This result suggests that the mid-IR nebular emission is a cleaner tracer of the relative contributions of AGN and star formation activity than optical, despite the fact that the optical \oiiirl\ line is a tracer of the accretion luminosity as good as mid-IR lines like \nev, \oiv,\ and \neiii\ (Sect. \ref{sec:lacc_tracers}). In a similar way, \cite{Abel2008} compare combined photoionization models of AGN and star formation with observations of optically classified starbursts and found that, the mid-IR is more sensitive to the presence of an AGN, even when the AGN contribution to the total luminosity (intended as luminosity of the incident radiation of the spectral models and not that of the mid-IR continuum in this specific case) is low ($\approx 0.01-0.1\%$).

Combined models are useful to determine the relative contribution from AGN and star formation to the total energy output of a galaxy \citep{Abel2008}. By means of only the nebular information of specific lines (see the relations above), one can already achieve important clues on the galaxy type and an order-of-magnitude estimate of the relative contributions from the AGN activity and star formation to the mid-IR emission in a given galaxy (Fig. \ref{fig:ratios_fagn}). This is because the line transitions that we find in the mid-IR regime have a wide range of ionization energies (Fig. \ref{fig:ionspec}), from ions that can be ionized (and excited) by both AGN and star formation to those (i.e., highest ionization potential) requiring the presence of a strong radiation field, as that from AGN or shocks. Another reason is that mid-IR emission traces the radiation from heavily dusty regions, which can be completely obscured at optical/UV wavelengths (Sect. \ref{sec:dust_caveats}). 

These ratio diagnostics will be instrumental for the design and interpretation of future mid-IR observations at higher redshift (e.g., PRIMA) as for example at the peak of the cosmic star formation history ($z\approx1-2$), where intense black hole accretion and starbursting events are expected to coexist in the same galaxy. Another reason for their relevance at high redshift is the fact that the classification via BPT diagrams of AGN and star-formation-dominated sources is less efficient at earlier cosmic times, in particular as results of the decreasing of the metal content with increasing $z$ \citep{Groves2006c, Feltre2016, Hirschmann2019}.

\subsection{Additional sources of ionizing radiation}\label{sec:additional_sources}

We find that pure AGN models alone do not reproduce the entire variety of mid-IR line ratios observed in our sample.
Given that our targets have a significant component of star formation, this is a natural source of additional ionizing photons to invoke. The addition of such a nebular emission component can explain the observations for the targets with low f$_{\rm AGN}$. As illustrated in Sect. \ref{sec:shock} and Fig. \ref{fig:mir_diagn}, these same data can be explained by adding to the AGN-driven photoionization, the emission from gas ionized by shocks from the ejecta of the AGN or galactic winds. The main issue here is to find a diagnostic enabling the identification of one or another source of ionization. This is not straightforward, because shocks are physically linked to black hole accretion and star formation phenomena, through AGN-driven outflows, galactic winds and the expansion of \ion{H}{ii} regions into the ISM for example. Galaxy mergers \citep[e.g.,][]{Guillard2012} and the gas infall within galaxies \citep[e.g.,][]{Vidal-Garcia2021,Lehmann2022} are also sources of shock-ionized emission. 

Among all the line ratio diagrams explored here, we find those involving \oi/\ha\ to best distinguish between AGN, star formation and shocks \citep[bottom-right panel of Fig. \ref{fig:mir_opt_diagn} and Fig. \ref{fig:mir_opt_appendixO1}; see also][]{Hirschmann2019, Kewley2019}. This ratio is a known tracer of shocks in neutral gas \citep{Allen2008, Ho2014, Riffel2020}, but we caution that the \oi\ emission may come from regions further from the central AGN compared to the central regions where high-ionization lines originate. In addition, when exploiting the combined diagrams in Sect. \ref{sec:mixed}, mid-IR and optical line ratios may probe different regions depending on dust attenuation. As discussed later in Sects. \ref{sec:dust_caveats} and \ref{sec:conclusions}, spatially resolved spectroscopy is crucial to address these issues.

Diagrams based only on mid-IR line ratios are less affected by dust attenuation and are promising diagnostics to disentangle AGN from star formation activity. As an example, young stars do not produce enough ionizing photons to triply ionize oxygen or quadruply ionize Ne, with stellar models predicting equivalent widths of the high-ionization  \oiv\ and \nev\ lines that are very weak or null. However, nebular emission in \hii\ regions can significantly contribute to the \neii\ emission, explaining the observed line ratios and making the \oiv/\neii\ versus \nev/\neii\ good AGN versus star formation diagnostics. While all the diagnostics shown in Fig. \ref{fig:mir_diagn} are auspicious to identify AGN from star formation, predictions from shock models overlap with either the AGN or star formation model grids depending on the combination of line ratios (see Fig. \ref{fig:mir_diagn_appendix}). This makes more complex the identification of the most effective diagrams to identify the dominant ionizing source among these three. An efficient way to quantify the goodness of the different diagnostics is that of modeling realistic populations of galaxies, as shown in \cite{Hirschmann2017, Hirschmann2019, Hirschmann2022}.

The FWHM values of the mid-IR lines are in the range of $\approx 400-1000$ km/s, typical of the narrow line in AGN and, at the same time, consistent with the velocities of fast-radiative shocks. Likely, an in depth study of the line profile in high-resolution mid-IR spectra could help identify targets whose line emission is dominated by shock-driven ionization, but this is outside the scope of this work. 

The most plausible hypothesis is that shock and star-formation-driven ionization bring significant contributions to different emission features observed in the spectra of our objects. This is likely because AGN, star formation activity and shocks all coexist in Seyfert galaxies. To unravel their relative importance, spatial information from spectroscopically resolved data as those from JWST/MIRI becomes crucial. 

\subsection{Caveats}\label{sec:caveats}

\subsubsection{Optical spectra flux calibration}\label{sec:oiii_caveats}

Obtaining an absolute flux calibration of the optical spectra was not possible with SALT data alone because of the variable pupil size of the telescope (typically by up to 20\%; see Sect. \ref{sec:salt}). In Sect. \ref{sec:oiii_fluxcalibration}, we appeal to published V-band aperture photometry for estimating the absolute flux of the \oiii\ line. Among the published measurements, we considered the 10$\arcsec$ diameter circular aperture photometry from \cite{Hunt1999b} to select the closest (even though not identical) area to that covered by the 1D SALT spectra extractions (performed for a slit length of 11$\arcsec$). 
Since the \oiii\ absolute flux estimates impact only Sects. \ref{sec:opt_Xray} and \ref{sec:opt_MIR} of this work and not the main results of the analysis (Sec. \ref{sec:model_comparison}), we refrain from performing new data analysis on archive images and spectra involving new aperture matching and photometry. However, we compared the absolute \oiii\ fluxes from SALT to those from the literature collected by \cite{Malkan2017} finding a good agreement, with $\approx70$\% of the targets within $1\sigma$ from the $1:1$ relation. In addition, the relations between the luminosities of mid-IR lines (\neiii, \nev,\ and \oiv) and the \oiii\ luminosity, either from SALT or \cite{Malkan2017}, are in remarkable agreement (left panels of Fig. \ref{fig:LoiiiLIR}). We, therefore, confirm that, within the accuracy needed for our purposes, our \oiii\ flux measurements are robust against absolute flux calibration.

\subsubsection{Dust attenuation and obscuration}\label{sec:dust_caveats}

Dust is a potential source of inaccuracy of our analysis as it impacts the observed line fluxes and ratios.
These are usually corrected by dust attenuation following different prescriptions, which are also source of uncertainty. Specifically, in Sect. \ref{sec:spectral_analysis} we derive dust attenuation, A$_{\rm V}$, from the Balmer decrement when \ha\ and \hb\ lines are both detected in the spectra, and consider A$_{\rm V}$ from SED fitting in the other cases.

\begin{figure}
\centering
  \includegraphics[width=9cm]{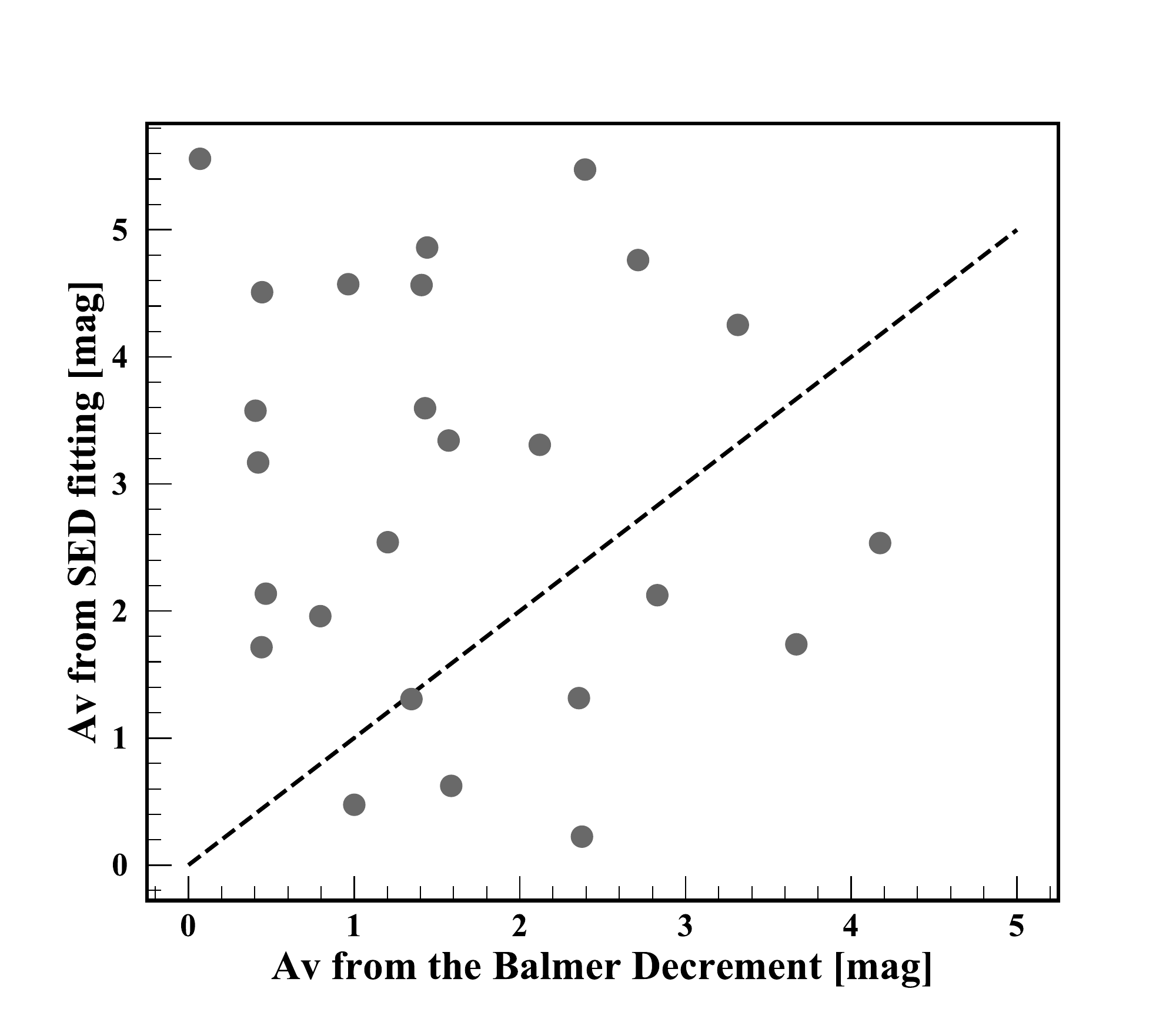}
     \caption{{Comparison between the V-band dust attenuation, A$_{\rm V}$, computed from the Balmer decrement ($x-$axis) and that from SED fitting ($y-$axis), as described in Sect. \ref{sec:line_measurements}.}}
     \label{fig:Av}
\end{figure}

We found a significant discrepancy in the values of A$_{\rm V}$ inferred from the two methods, with a large scatter around the 1:1 line. This is shown in Fig. \ref{fig:Av} for the 25 objects were was possible to apply the Balmer decrement. This difference is not surprising as the two methods may probe different phases of the medium within our sources. The Balmer decrement probes the excited gas in the photoionized regions, but does not trace the absorption of ionizing photons by dust in \ion{H}{ii} regions before they ionize hydrogen (e.g., \citealp{Mathis1986}; see also \citealp{Charlot2002}). The dust attenuation from the SED fitting is inferred from the continuum emission as seen from broadband photometry and can include the contribution from the diffuse gas in the host galaxy. This has been found to be lower than that obtained from the Balmer decrement in star-forming galaxies \citep{Wild2011, Price2014}. The SED fitting procedure used in G16 to infer dust attenuation assumes the two-component model by \cite{Charlot2000}, where the radiation is attenuated by diffuse ISM and dust in the birth clouds. Therefore, we primary asses the scatter in Fig. \ref{fig:Av} to uncertainties and degeneracies of the two methods. 

However, as already mentioned above, the impact of the A$_{\rm V}$ is relatively small for optical line ratios. For reference, by adopting a \cite{Cardelli1989} attenuation curve as in Sect. \ref{sec:line_measurements}, an A$_{\rm V}$ of 1(4) would affect the \oiii/\hb, \nii/\ha, \sii/\ha, and \oi/\ha\ line ratios of 0.1(0.4), 0.007(0.03), 0.06(0.2), and 0.1(0.4) dex, respectively.

When presenting combined optical and mid-IR line-ratios diagrams in Sect. \ref{sec:mixed}, we assume that optical and mid-IR lines come from the same regions. However, we cannot determine whether a certain fraction of the mid-IR flux probes heavily obscured regions where optical lines are not detected because of dust obscuration. In the case in which the UV/optical lines (and, hence, the Balmer decrement) only trace unattenuated gas emission, the intrinsic optical line ratios shown in Fig. \ref{fig:mir_opt_diagn} would move toward the areas of the diagrams populated by AGN or star formation, depending on whether the heavily obscured areas are the most central regions or star-forming regions. In this case, one would need to appeal to mid-IR line ratios to have a more secure determination of the contribution from the obscured and heavily obscured (i.e., not detectable via optical/UV lines) sources of ionizing radiation. 

We would like to stress that information from detailed kinematic studies and integral field spectroscopy is crucial for tackling the caveats discussed in this section. Combining spatially resolved data in the optical (e.g., from MUSE on the VLT) with those in the mid-IR from JWST/MIRI will ensure the same regions of the galaxies are covered. At the same time, the comparison of the dust maps obtained from the Balmer ratios of optical lines with maps of mid-IR line ratios will enable the identification of heavily obscured regions and a more detailed investigation of the impact of dust on the diagnostic diagrams proposed in this work.

\section{Summary and conclusions}\label{sec:conclusions}

We collected and analyzed a set of homogeneous optical spectra of 42 local Seyfert galaxies observed by SALT and interpreted the optical line ratios measured on these spectra along with mid-IR ratios available from the literature through the exploitation of line emission models of different ionizing sources. In what follows, we summarize the main results of this work and highlight future perspectives:

\begin{itemize}

\item[$\bullet$] Measurements of the \oiii\ line intensities after absolute flux calibration are consistent with previous measurements of the relation between \oiii\ luminosity and X-ray luminosities (Sect. \ref{sec:opt_Xray}) and with other tracers (\nev, \oiv,\ and \neiii\ mid-IR lines) of bolometric luminosities (Sect. \ref{sec:opt_MIR}).

\item[$\bullet$] We compared the \oiii\ line intensity with that of the \nev, \oiv,\ and \neiii\ mid-IR lines, which are good tracers of the AGN accretion luminosities.\ We provide the relations between \oiii\ and the mid-IR lines so that they can serve for the design of future observations (Sect. \ref{sec:opt_MIR}). 

\item[$\bullet$] When comparing observations with a suite of pure AGN photoionization models of the NLRs, we find that mid-IR line ratios of sources with high f$_{\rm AGN}$ (i.e., $>40\%$) are well reproduced (Sect. \ref{sec:AGNcomparison}). 

\item[$\bullet$] The AGN photoionization models presented in Sect. \ref{sec:photomod} struggle to reproduce the low f$_{\rm AGN}$ (i.e., $<40\%$) tail of targets. We explored a wider parameter space, different shapes of the AGN incident radiation, and calculations with different spatial extents in Sect. \ref{sec:agnmodels}. We find that even using extreme values of the power law index, $\alpha=-0.9$ (i.e., harder ionizing spectrum), we can only partially reproduce the mid-IR ratios of the Seyferts with low f$_{\rm AGN}$.

\item[$\bullet$]The exploration of other sources of ionizing photons, in addition to the AGN, suggests that either a contribution from star formation or a shock can help explain the observed mid-IR ratios of the low f$_{\rm AGN}$ targets (Sects. \ref{sec:sf} and \ref{sec:shock}). 

\item[$\bullet$] While optical line ratios do not show any dependence on the fractional contribution of the AGN to the $5-40$ \mum\ mid-IR continuum f$_{\rm AGN}$, some specific mid-IR line ratios, namely \nev/\neii, \oiv/\neii, \neii/\neii, and \siv/\neii,\ increase with f$_{\rm AGN}$, resulting in much cleaner tracers of the AGN contribution to the total SED. We report relations between f$_{\rm AGN}$ and the line ratios of interest in Sect.\ref{sec:fagn}.  

\item[$\bullet$]Combined optical and mid-IR diagrams based on the \oi/\ha\ ratio versus mid-IR ratios (Sect. \ref{sec:mixed} and Fig. \ref{fig:mir_opt_appendixO1}) are the most promising as a diagnostic of the three ionizing sources: AGN, star formation, and shocks. This is based on the assumption that the nebular lines originated in the same region. Diagnostic diagrams involving only mid-IR lines would be less affected by these problems; however, the grid of models of AGN, shocks, and star-forming galaxies is less clearly separated in the diagrams in Fig. \ref{fig:mir_diagn_appendix} than in, for example, the \oi/\ha\ versus \oiv/\neiii\ diagrams shown in Fig. \ref{fig:mir_opt_appendix}. 

\end{itemize}

A better way to  quantify how effective mid-IR and combined optical and mid-IR diagnostics are in distinguishing between the different ionizing sources is by modeling the emission of realistic populations of galaxies. A way to achieve this is to combine AGN, shocks, and star formation in a physically motivated way, inferring, for example, the information from cosmological hydrodynamical simulations, similar to the method presented in \cite{Hirschmann2017, Hirschmann2019, Hirschmann2022}.

Given that the observations of our targets with a contribution from the AGN to the mid-IR continuum higher than $\approx40\%$ can be explained by AGN photoionization models with a range of spectral slopes, $\alpha$, we propose combining optical and mid-IR information with analysis tools based on advanced statistical techniques, such as \textsc{BEAGLE} \citep{Chevallard&Charlot2016,Vidal-Garcia2022}, and exploring how best these combined observables can constrain the steepness of the ionizing radiation that cannot be well constrained by using the optical information alone. 
\textsc{BEAGLE}, which already incorporates the nebular emission due to star formation using the \cite{Gutkin2016} models, has recently been updated with the inclusion of the AGN photoionization models of F16 and can now combine the nebular emission from the AGN and \hii\ regions for interpreting the emission line spectra of Type 2 AGN (\textsc{BEAGLE-AGN};  \citealt{Vidal-Garcia2022}). Analyzing the goodness of \textsc{BEAGLE-AGN} Bayesian fits on optical and mid-IR line ratios using composite models of AGN and star-formation-driven ionization would also help identify the need of a contribution from additional sources of ionizing photons, such as fast radiative shocks.

Regarding this last point, mid-IR spatially resolved spectroscopy such as that from JWST/MIRI will be instrumental in identifying the regions of the host galaxy affected by shocks and in studying the influence of the AGN-driven ionization on galactic scales. We note that our sample is ideal for identifying sources to be targeted by integral field unit spectrographs and that MIRI observations of the Seyfert NGC\,7469 in our sample are already planned within the Early Release Science program 1328 (PI. Armus L).

The identification of mid-IR line ratios that trace AGN activity are important for the design of observations with current \cite[e.g., JWST/MIRI; see also][]{Satyapal2021, Richardson2022} and future facilities (e.g., PRIMA).
 Future mid-IR observations will expand current studies in the local Universe toward the peak of SFR densities ($z\approx1-2$) and beyond. The models presented here can serve as preparatory work for the design of these future missions, and they can easily be coupled with cosmological simulations, following for example the approach of \cite{Hirschmann2017,Hirschmann2019}, or incorporated into phenomenological simulation tools such as the Spectro-Photometric Realisations of IR-Selected Targets at all $z$ \citep[SPRITZ;][]{Bisigello2021}, 

To conclude, within the  SALT Spectroscopic Survey of IR 12MGS Seyfert Galaxies we have obtained SALT spectra that cover a shorter wavelength range (down to 3600 \AA) than those presented here to cover the [\ion{O}{ii}]$\lambda3726,3729$ and [\ion{Ne}{iii}]$\lambda3869$ lines and obtain \hb\ line measurements for all 42 targets in the sample. This will enable the explorations of diagnostics other than those used to identify the dominant ionizing source, such as metallicity and density diagnostics, and allow us to compare them with the less calibrated mid-IR diagnostics. Investigating and identifying which mid-IR ratios can best trace the physical conditions of the ionizing gas is crucial for driving the design of future mid-IR spectrographs. 

\begin{acknowledgements}
        AF and CG acknowledge support from grant PRIN MIUR 2017 20173ML3WW. AVG acknowledges support from the European Research Council Advanced Grant MIST (No 742719, PI: E. Falgarone). 
        AM acknowledges support from the National Research Foundation of South Africa (Grant Numbers 110816 and 132016), and financial support from the Swedish International Development Cooperation Agency (SIDA) through the International Science Programme (ISP) - Uppsala University to the University of Rwanda through the Rwanda Astrophysics, Space and Climate Science Research Group (RASCSRG).
         PV and ERC acknowledges support from the National Research Foundation of South Africa.
        FS acknowledges financial support from the Italian Ministry of University and Research - Project Proposal CIR01$\_$00010, and funding from the INAF mainstream 2018 program ”Gas-DustPedia: A definitive view of the ISM in the Local Universe“.
        JC acknowledges funding from the ERC Advanced Grant 789056 “FirstGalaxies” (under the European Union’s Horizon 2020 research and innovation programme).
        ECL acknowledges support of an STFC Webb Fellowship (ST/W001438/1).
        OLD is grateful to the Fundação de Amparo à Pesquisa do Estado de São Paulo  (FAPESP) and to Conselho Nacional de Desenvolvimento Científico e Tecnológico (CNPq) for the financial support.
        SM acknowledges funding from the INAF ``Progetti di Ricerca di Rilevante Interesse Nazionale'' (PRIN), Bando 2019 (project: ``Piercing through the clouds: a multiwavelength study of obscured accretion in nearby supermassive black holes'’). LM and MV acknowledge financial support from the Inter-University Institute for Data Intensive Astronomy (IDIA), a partnership of the University of Cape Town, the University of Pretoria, the University of the Western Cape and the South African Radio Astronomy Observatory, and from the South African Department of Science and Innovation's National Research Foundation under the ISARP RADIOSKY2020 Joint Research Scheme (DSI-NRF Grant Number 113121) and the CSUR HIPPO Project (DSI-NRF Grant Number 121291).

  This work made use several open source \textsc{python} packages: \textsc{numpy} \citep{vanderWalt2011}, \textsc{pyspekit} \citep{Ginsburg2011},  \textsc{astropy}, a community-developed core Python package for Astronomy \citep{astropy:2013, astropy:2018, astropy:2022}, \textsc{PyNeb} \citep{Luridiana2015} and \textsc{scipy} \citep{Virtanen2020}. 

\end{acknowledgements}

% WARNING
%-------------------------------------------------------------------
% Please note that we have included the references to the file aa.dem in
% order to compile it, but we ask you to:
%
% - use BibTeX with the regular commands:
%   \bibliographystyle{aa} % style aa.bst
%   \bibliography{Yourfile} % your references Yourfile.bib
%
% - join the .bib files when you upload your source files
%-------------------------------------------------------------------

\bibliographystyle{aa}
\bibliography{main}

\begin{appendix}

\section{Diagnostic diagrams} \label{app:midIR_full}

\begin{figure*}
\centering
  \includegraphics[width=17cm]{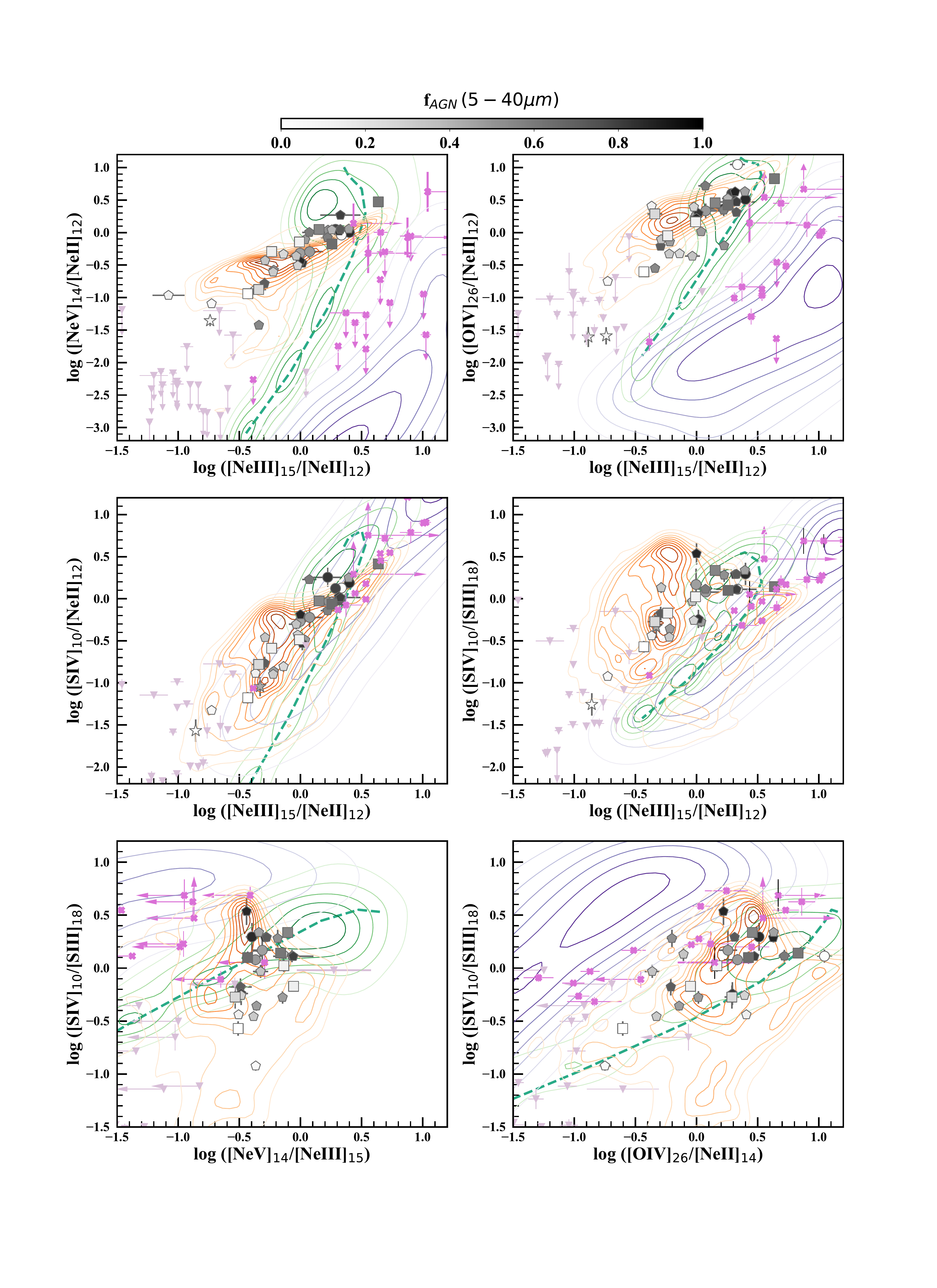}
     \caption{Same as Fig. \ref{fig:mir_diagn} but showing the SF and shock models with purple and orange contours, respectively. Pink crosses and downward pointing lilac triangles are observations of star-forming and dwarf galaxies, respectively. The former are from \cite{Bernard-Salas2009} and \cite{GouldingAlexander2009}, the latter from \cite{Cormier2015}.}
     \label{fig:mir_diagn_appendix}
\end{figure*}

\begin{figure*}
\centering
  \includegraphics[width=17cm]{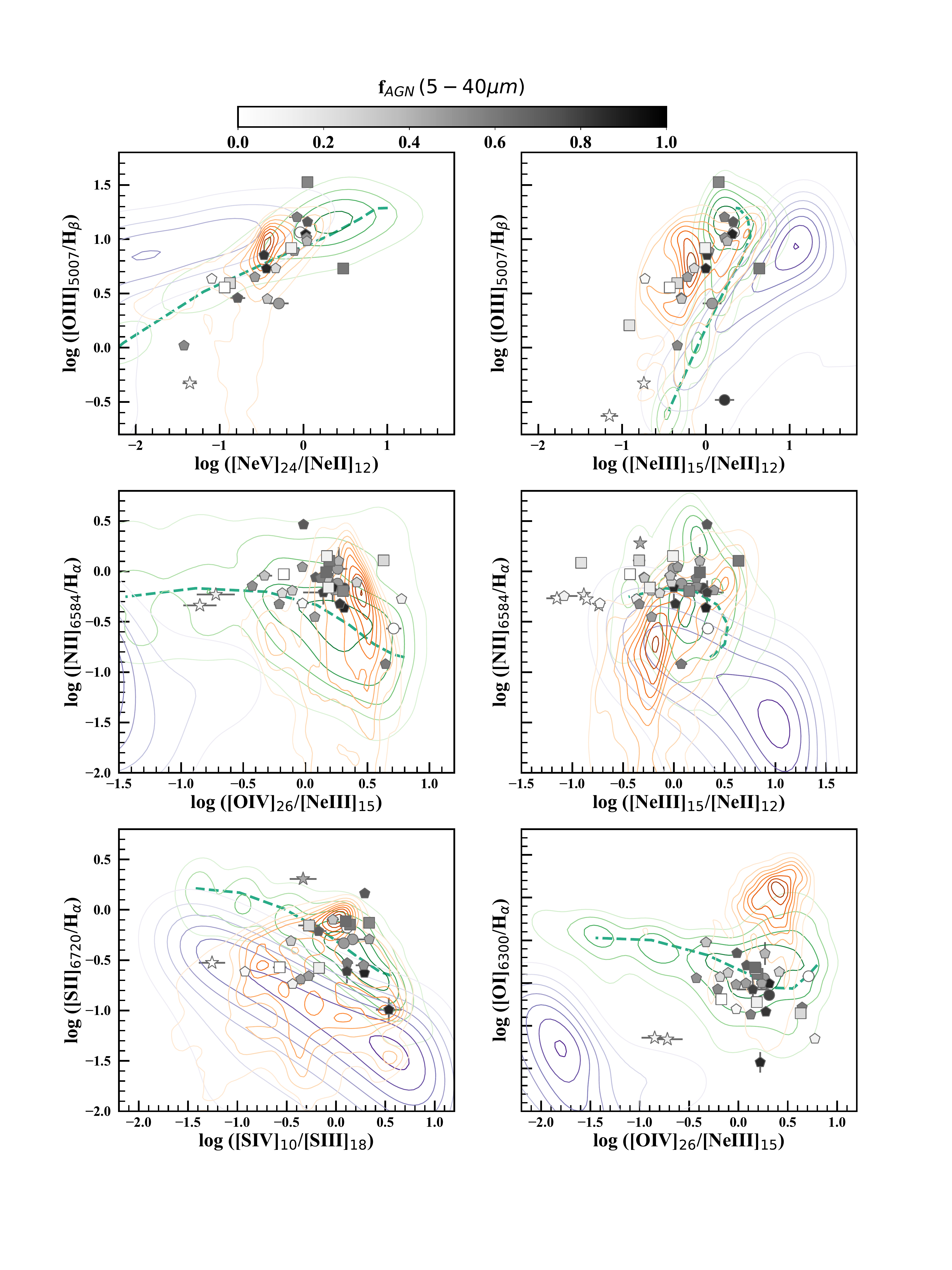}
     \caption{Same as Fig. \ref{fig:mir_opt_diagn} but showing the SF and shock models with purple and orange contours, respectively.}
     \label{fig:mir_opt_appendix}
\end{figure*}

\begin{figure*}
\centering
  \includegraphics[width=17cm]{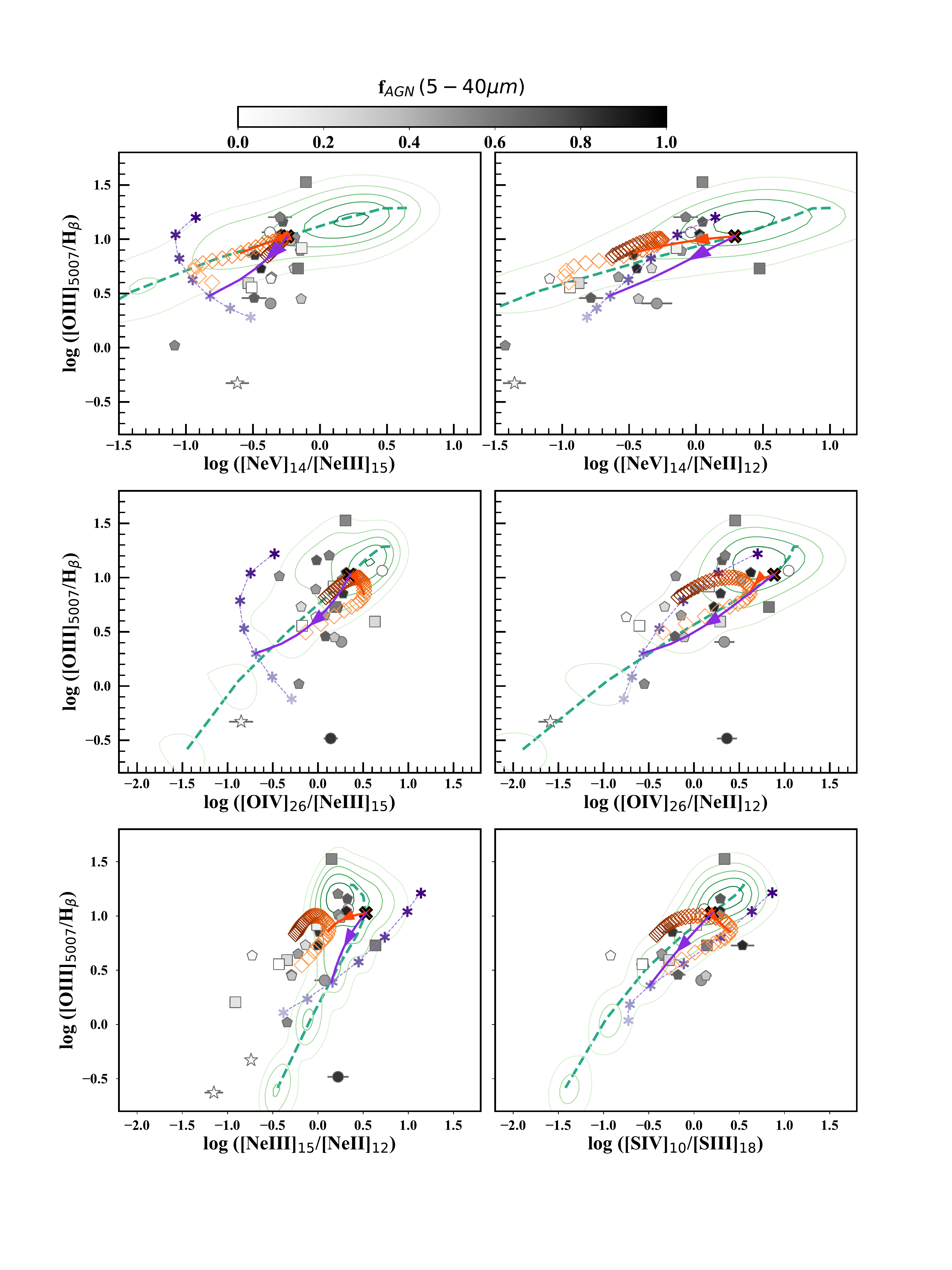}
     \caption{Examples of \oiii/\hb\ versus mid-IR ratio diagrams. Symbols and colors of the observations and models are the same as in Fig. \ref{fig:mir_diagn}.}
     \label{fig:mir_opt_appendixO3}
\end{figure*}

\begin{figure*}
\centering
  \includegraphics[width=17cm]{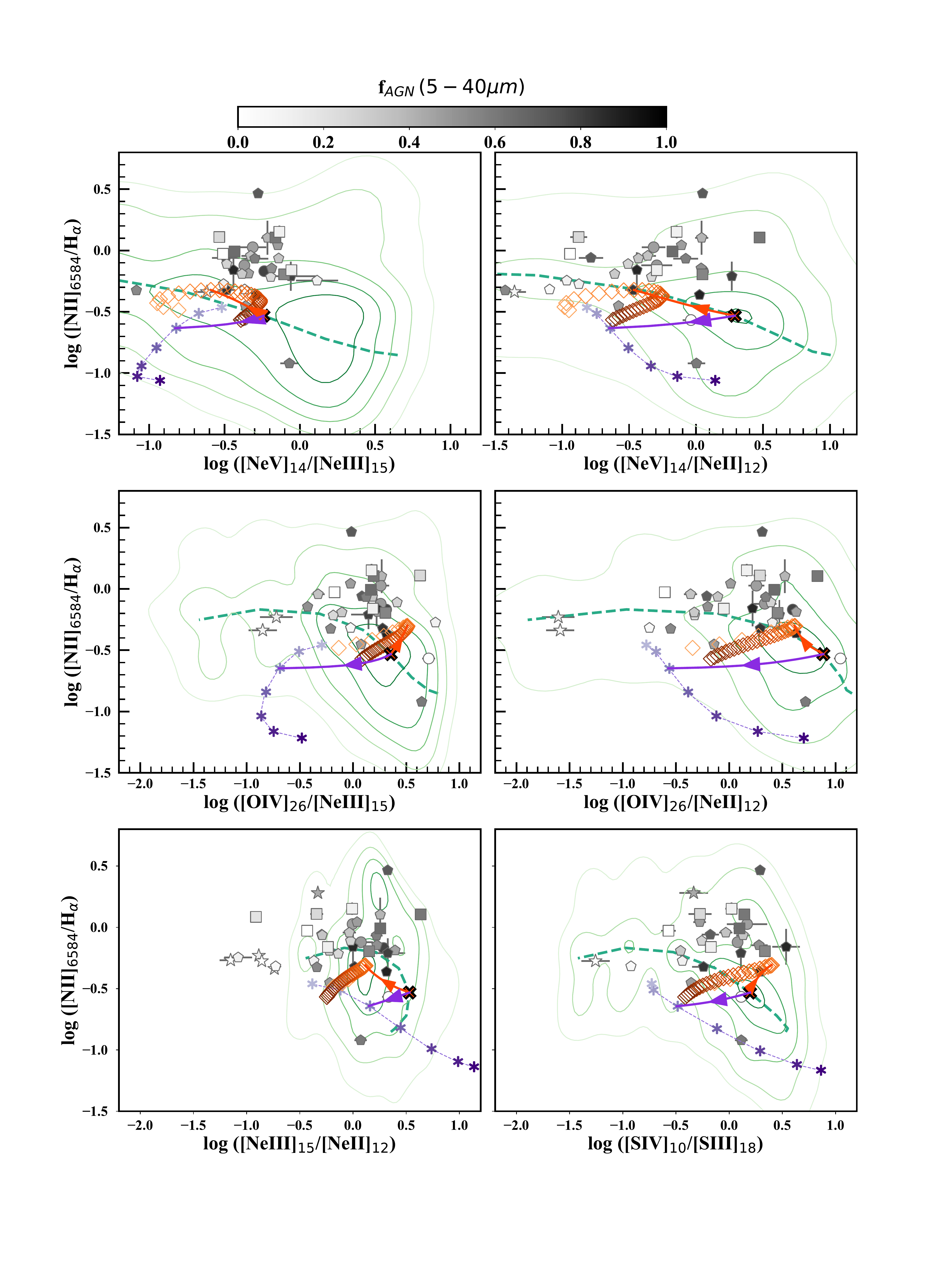}
     \caption{Examples of \nii/\ha\ versus a mid-IR ratio diagrams. Symbols and colors of the observations and models are the same as in Fig. \ref{fig:mir_diagn}.}
     \label{fig:mir_opt_appendixN2}
\end{figure*}

\begin{figure*}
\centering
  \includegraphics[width=17cm]{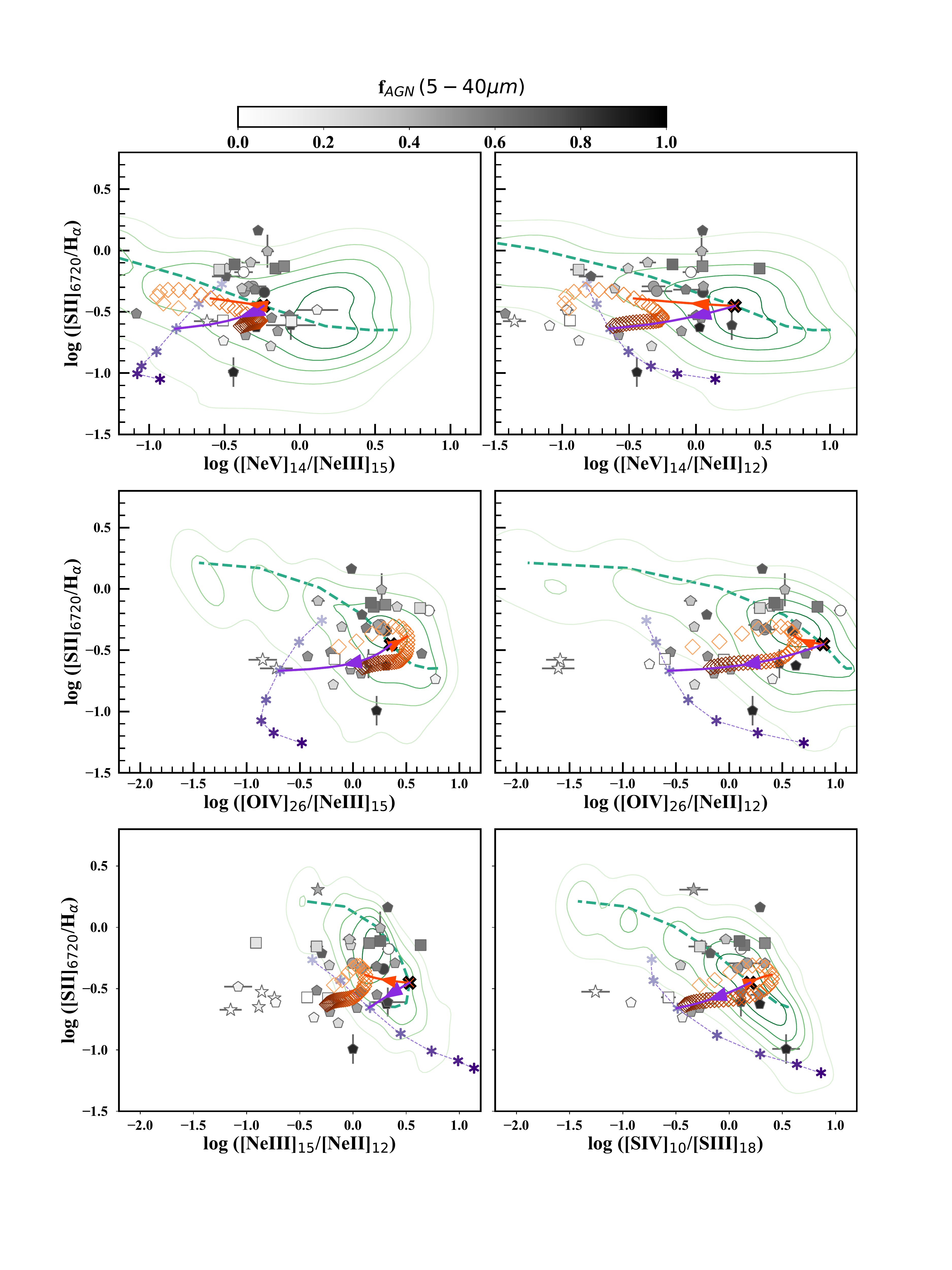}
     \caption{Examples of \sii/\ha\ versus mid-IR ratio diagrams. Symbols and colors of the observations and models are the same as in Fig. \ref{fig:mir_diagn}.}
     \label{fig:mir_opt_appendixS2}
\end{figure*}

\begin{figure*}
\centering
  \includegraphics[width=17cm]{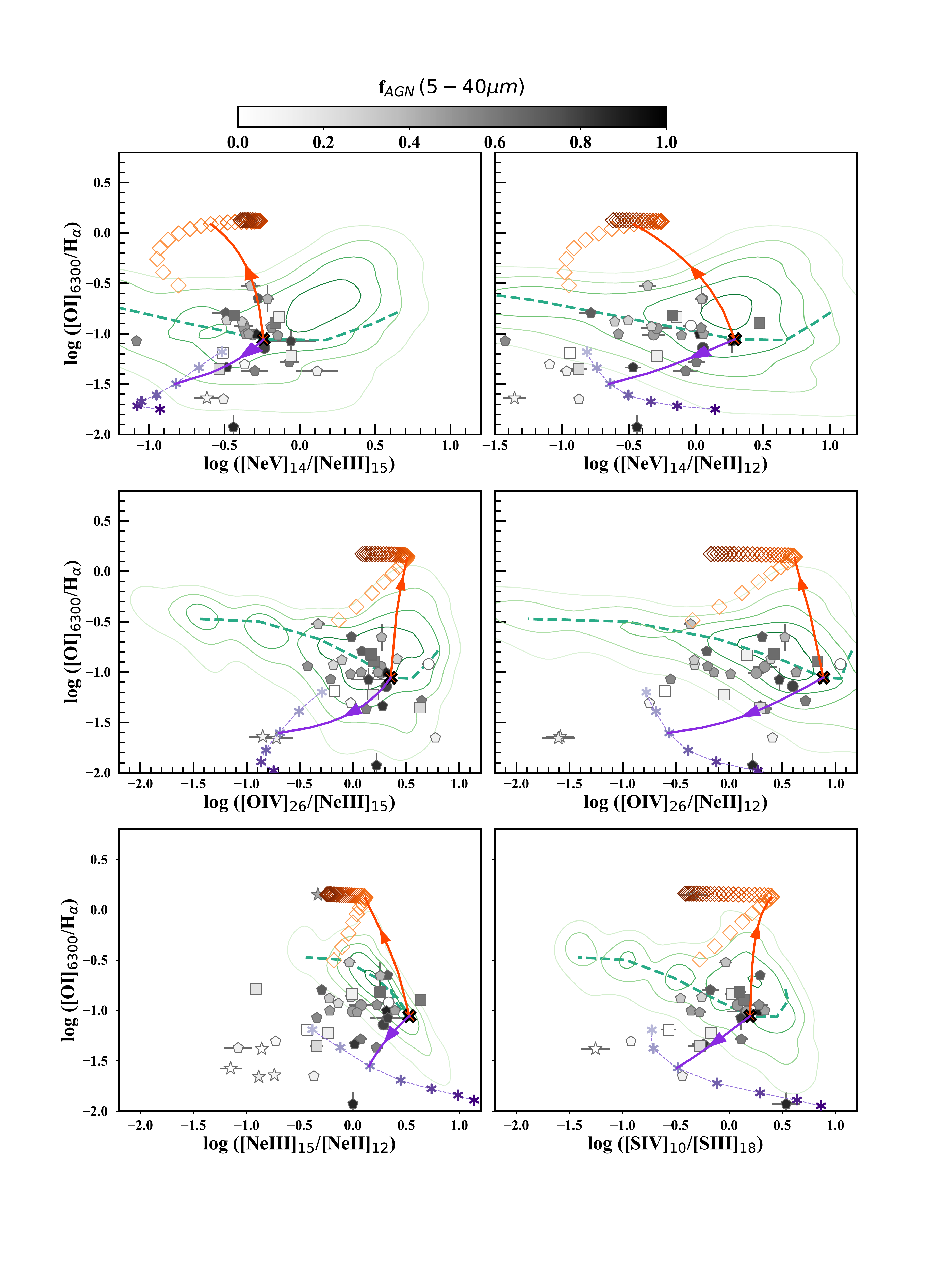}
     \caption{Examples of \oi/\ha\ versus mid-IR ratio diagrams. Symbols and colors of the observations and models are the same as in Fig. \ref{fig:mir_diagn}.}
     \label{fig:mir_opt_appendixO1}
\end{figure*}

The diagnostic diagrams in Figs. \ref{fig:mir_diagn_appendix} to \ref{fig:mir_opt_appendixO1} are an extension of the diagrams shown in Fig. \ref{fig:mir_diagn} and \ref{fig:mir_opt_diagn} of the main text.
In particular, the colored contours in Figs. \ref{fig:mir_diagn_appendix} and \ref{fig:mir_opt_appendix} show the entire grids of the SF and shock models with $Z<1/3\,Z_{\odot}$. In addition, in Fig. \ref{fig:mir_diagn_appendix} we show data points for star-forming galaxies \citep{Bernard-Salas2009, GouldingAlexander2009}  and dwarf galaxies  \cite{Cormier2015}, respectively, collected by \cite{Fernandez-Ontiveros2016}. These starburst galaxies have \nev/\neiii\ $<0.1$, lower than most of the values of our targets \citep[see also figure 11 of][]{Groves2006b}. We note that the SF models with ionization parameter lower than the values covered by the \cite{Gutkin2016} model grid can explain the line ratios observed in star-forming galaxies (down-pointing triangles in Fig. \ref{fig:mir_diagn_appendix}), as shown in  \cite{Fernandez-Ontiveros2016}. 

Figures \ref{fig:mir_opt_appendixO3} to \ref{fig:mir_opt_appendixO1} show all the possible combinations of optical and mid-IR line ratios, while Fig. \ref{fig:mir_opt_diagn} reports some illustrative examples. 

\end{appendix}

\end{document}